\documentclass[conference]{IEEEtran}
\IEEEoverridecommandlockouts
\usepackage{cite}
\usepackage{amsmath,amssymb,amsfonts}
\usepackage{algorithmic}
\usepackage{graphicx}
\usepackage{textcomp}
\usepackage[table]{xcolor}
\usepackage{arydshln}
\usepackage{booktabs}
\usepackage{graphbox}
\usepackage{url}
\usepackage{subfig}
\usepackage{comment}
\usepackage{hyperref}
\def\BibTeX{{\rm B\kern-.05em{\sc i\kern-.025em b}\kern-.08em
    T\kern-.1667em\lower.7ex\hbox{E}\kern-.125emX}}

\begin{document}

\title{
% Performance evaluation of point cloud objective quality metrics \\
% Large-scale performance evaluation of objective quality metrics on point clouds
Assessing objective quality metrics for JPEG and MPEG point cloud coding
    
% \thanks{The authors would like to acknowledge support from the Swiss National Scientific Research project entitled ``Compression of Visual information for Humans and Machines (CoViHM)” under grant number 200020\_207918.}

\thanks{The authors acknowledge support from the Swiss National Scientific Research project under grant number 200020\_207918.}

% \thanks{Acknowledgements have been removed to respect the double-blind policy of the QoMEX 2024 conference}

}

% \begin{comment}
\author{\IEEEauthorblockN{Davi Lazzarotto, Michela Testolina, Touradj Ebrahimi}
\IEEEauthorblockA{Multimedia Signal Processing Group (MMSPG)\\
École Polytechnique Fédérale de Lausanne (EPFL)\\
Lausanne, Switzerland \\
davi.nachtigalllazzarotto@epfl.ch, michela.testolina@epfl.ch, touradj.ebrahimi@epfl.ch}
% \and
% \and
}
% \end{comment}

% \author{\\[2.0ex]
% \IEEEauthorblockN{Anonymous QoMEX 2024 submission}\\[2.0ex]}

\maketitle

\begin{abstract}

As applications using immersive media gained increased attention from both academia and industry, research in the field of point cloud compression has greatly intensified in recent years, leading to the development of the MPEG compression standards V-PCC and G-PCC, as well as the more recent JPEG Pleno learning-based point cloud coding. 
Each of the standards mentioned above is based on a different algorithm, introducing distinct types of degradation that may impair the quality of experience when lossy compression is applied. 
Although the impact on perceptual quality can be accurately evaluated during subjective quality assessment experiments, objective quality metrics also predict the visually perceived quality and provide similarity scores without human intervention. 
Nevertheless, their accuracy can be susceptible to the characteristics of the evaluated media as well as to the type and intensity of the added distortion. 
While the performance of multiple state-of-the-art objective quality metrics has already been evaluated through their correlation with subjective scores obtained in the presence of artifacts produced by the MPEG standards, no study has evaluated how metrics perform with the more recent JPEG Pleno point cloud coding. 
In this paper, a study is conducted to benchmark the performance of a large set of objective quality metrics in a subjective dataset including distortions produced by  JPEG and MPEG codecs. 
The dataset also contains three different trade-offs between color and geometry compression for each codec, adding another dimension to the analysis. 
Performance indexes are computed over the entire dataset but also after splitting according to the codec and to the original model, resulting in detailed insights about the overall performance of each visual quality predictor as well as their cross-content and cross-codec generalization ability.

\end{abstract}

\begin{IEEEkeywords}
Point cloud, Quality assessment, Objective metrics, Benchmarking
\end{IEEEkeywords}

\section{Introduction}

% \begin{figure}

%     \centering
%     % \begin{minipage}[b]{0.95\linewidth}

%     \centering
%     \subfloat[\emph{Bouquet}]{
%     \begin{minipage}[b]{0.32\linewidth}
%     \centering
%     \includegraphics[width=\linewidth]{Figures/Dataset/Bouquet.png}
%     \end{minipage}}
%     \subfloat[\emph{Soldier}]{
%     \begin{minipage}[b]{0.32\linewidth}
%     \centering
%     \includegraphics[width=\linewidth]{Figures/Dataset/Soldier.png}
%     \end{minipage}}
%     \subfloat[\emph{Boxer}]{
%     \begin{minipage}[b]{0.32\linewidth}
%     \centering
%     \includegraphics[width=\linewidth]{Figures/Dataset/Boxer.png}
%     \end{minipage}}

%     \subfloat[\emph{StMichael}]{
%     \begin{minipage}[b]{0.32\linewidth}
%     \centering
%     \includegraphics[width=\linewidth]{Figures/Dataset/StMichael.png}
%     \end{minipage}}
%     \subfloat[\emph{Thaidancer}]{
%     \begin{minipage}[b]{0.32\linewidth}
%     \centering
%     \includegraphics[width=\linewidth]{Figures/Dataset/Thaidancer.png}
%     \end{minipage}}
%     \subfloat[\emph{House\_without\_roof}]{
%     \begin{minipage}[b]{0.32\linewidth}
%     \centering
%     \includegraphics[width=\linewidth]{Figures/Dataset/House_without_roof.png}
%     \end{minipage}}

%     % \end{minipage}
    
%     \caption{Point clouds employed in this assessment study.}
%     \label{fig:dataset}
% \end{figure}

The recent rise of immersive applications such as virtual and augmented reality has highlighted the need for effective representations of three-dimensional (3D) models with high visual quality. 
% Although meshes have been extensively used in the field of computer graphics for the representation of content produced by humans, 
Point clouds have gained popularity for the representation of real-world objects and scenes as a set of unconnected points with associated attributes such as colors. 
Since they can contain very large amounts of data, usually requiring millions of points to represent objects with an acceptable level of detail, 
effective compression algorithms must be available to reduce the burden on transmission and storage. %, with however the undesired side effect of introducing visual distortion. 

This need led standardization groups such as JPEG and MPEG to create activities that culminated in the development of three compression standards. 
The MPEG Geometry-based Point Cloud Compression (G-PCC) is based on the octree representation and encodes both geometry and color attributes directly in the 3D domain with handcrafted transforms. 
The Video-based Point Cloud Compression (V-PCC) splits the point cloud into patches that are separately projected onto different views, producing depth and color maps that are packed together and encoded with state-of-the-art video codecs. 
More recently, JPEG Pleno has started the development of a compression standard based on deep learning techniques, separately encoding geometry in the 3D domain and color as projected maps using JPEG AI. 

Each of these three approaches behaves differently according to the characteristics of the encoded model and produces distinct types of visual artifacts. 
While point clouds compressed with G-PCC have their density reduced, the artifacts of V-PCC usually follow the behavior of the video codec used to encode the 2D maps. 
JPEG Pleno also encodes color in the 2D domain, and therefore its effect is similar to JPEG AI, but the autoencoders used to compress the coordinates can add irregular artifacts such as holes in some regions and increased point density in others. 
The impact of such artifacts on the perceptual quality of the point clouds can only be reliably evaluated during subjective quality experiments, which are expensive and time-consuming. 
% through experiments where subjects visualize decoded models at different compression rates and are asked to rate them. 
% However, subjective experiments can be expensive to organize, usually needing special settings only available in laboratories or a large pool of subjects if it is conducted using crowdsourcing. 
For this reason, objective quality metrics can also be used to simulate the human visual system and produce quality scores based only on computations applied to the point cloud models.
% designed to have a high correlation with subjective perception based only on computations applied to the point cloud models. 
The performance of these metrics is often evaluated in benchmarking studies where the correlation between objective scores computed over a set of distorted point clouds and subjective mean opinion scores (MOS) is obtained. 
% the correlation between objective and  scores is obtained. 

In the recent literature, the performance of state-of-the-art point cloud objective quality metrics has been evaluated using datasets compressed with G-PCC and V-PCC. %, as well as other types of distortion unrelated to compression. 
However, artifacts generated by the more recent JPEG Pleno standard have not yet been considered for benchmarking. 
Moreover, in many of the previous studies, all reference models had similar density and voxelization precision, failing to consider how such parameters could affect quality predictions. % how the metrics handle distortions for those characteristics. 
Finally, recent promising metrics have not been independently evaluated and compared to more traditional algorithms. 

In this paper, a study is conducted to evaluate the performance of a large number of objective quality metrics using the MJ-PCCD dataset \cite{lazzarotto2024subjective}, which contains a set of distorted point clouds generated from six original models 
% selected from the JPEG Pleno test set \cite{JPEG-Pleno-PC-CTTC} 
using G-PCC, V-PCC, and JPEG Pleno. 
In addition to four bitrates defined separately for each codec, the dataset also contains three different trade-offs between color and geometry quality for each rate and codec. 
As a result, the correlation between objective and subjective scores of a particular metric is penalized if it does not properly balance the impact of geometric and color artifacts on perceptual quality, which would otherwise remain unnoticed when the same compression parameters are maintained throughout the dataset for each rate. 
Moreover, this dataset contains two point clouds where the point density is considerably lower than the rest, enabling the evaluation of the impact of sparsity on metrics performance. 
A large selection of 28 metrics separated into six categories are evaluated using the entire dataset as well as subsets obtained after splitting according to codec or content, producing numerous insights about their ability to model the human visual system in different circumstances.

\section{Related work}

\subsection{Objective quality assessment}

Numerous objective quality metrics for point clouds have been proposed in recent years, being mainly classified into three main categories: full reference (FR), no-reference (NR), and reduced reference (RR). 
FR metrics compare the distorted point cloud against a pristine reference, while NR metrics produce a final score based on the distorted model only. 
RR metrics can be viewed as an intermediary category, using only a limited set of features from the reference instead of the entire point cloud. 

% In the category of FR point cloud objective quality metrics, early studies used the point-to-point distance as the quality estimator, computing the distance of each point in the reference to its nearest neighbor in the distorted model and vice-versa, producing a final symmetric score through the mean square error or Hausdorff distance. % Reduce this sentence. Maybe merge this and the next two
% A variant of this metric was later proposed \cite{tian2017geometric} by computing the distance only on the direction of the normal vector. 
% For both metrics, the peak signal-to-noise ratio (PSNR) can be employed to provide the similarity between both models using the bounding box size to obtain the maximum distance used for its computation. 
% These metrics can be further classified as geometry-based since they do not consider the color values in their computation, differing from color-based metrics that compute a final score mainly based on the color values. 
% The first metric from the latter category is the color PSNR, which computes the difference between color values between nearest neighbors usually in the YUV colorspace. 
% However, unlike geometry-based metrics, color PSNR can still be implicitly affected by geometric distortions since operations such as point displacement or downsampling change the nearest neighbor computation. 

In the category of FR quality assessment, geometry-based metrics consider only geometric distortions while ignoring any other attributes. 
Early predictors included the point-to-point distance and the point-to-plane distance \cite{tian2017geometric}. % where only the displacement over the normal vector is considered. 
Later metrics take advantage of more complex geometric features such as the point-to-distribution \cite{Javaheri2020}, which computes a quality score based on the Mahalanobis distance, and the multiscale potential energy discrepancy (MPED) \cite{yang2022mped}, which measures the difference in point potential energy at multiple scales. %, a feature inspired on classical physics. 
The density-to-density distance \cite{m61195} has also been proposed targeting distortions produced by learning-based compression methods.

Color-based metrics produce a score computed only from color attributes, which can however still be implicitly affected by geometric distortions due to their impact on neighborhood association between point clouds. 
The color Peak Signal-to-Noise Ratio (PSNR) adapts the popular image metric to be computed in the 3D domain. 
% computes the average squared difference between color values from nearest neighbors usually in the YUV colorspace. 
Another metric computing the distance between color histograms of point clouds has been proposed in \cite{viola2020color}. 
GraphSIM \cite{yang2020inferring} employed graphs to compute structural similarities around key points of the point cloud, 
%using three moments of color gradients. 
while PointSSIM \cite{Alexiou2020a} calculated the structural similarity between point clouds using statistical estimators applied to local attribute distributions. % the distribution of attributes in local neighborhoods. %, where the variance of the luminance has been found as the best predictor. 
MS-GraphSIM \cite{zhang2021ms} and MS-PointSSIM \cite{lazzarotto2023towards} were later proposed to expand them to multiple scales. %incorporate metric values computed at multiple scales into one final score to improve their performance. 

Other metrics combine both geometry and color-based features. 
The joint point-to-distribution \cite{javaheri2021point} adapts the geometry-based metric\cite{javaheri2020point} to color attributes and combines them together into a single score. 
PCQM \cite{meynet2020pcqm} defines multiple geometry-based and color-based features, using logistic regression to find the linear combination optimizing the correlation. 

% While FR metrics employ the entire reference model to estimate the quality of a distorted model, RR methods were proposed using only a set of features from the reference. 
From the RR category, 
PCM\_RR \cite{Viola2020b} defines multiple features %based on the geometry, luminance, and normal vectors, 
and combines them through linear optimization,  
while RR-CAP \cite{zhou2023reduced} uses saliency-based features from the projection of the point cloud to compute the objective quality of the point cloud. 
Recently, many learning-based NR metrics have also been proposed. 
% NR metrics go even further and only need the distorted model, often using deep learning techniques to extract perceptual features. 
MM-PCQA \cite{zhang2022mm} fuses features obtained both from the point domain and projected images, % with symmetric cross-modality attention, 
while VQA-PC \cite{zhang2023evaluating} evaluates the quality using videos obtained from simulated camera paths. % captured from multiple camera positions.  

In addition to quality predictors designed specifically for point clouds, previous studies employed image quality metrics on projections of the point clouds. % at different angles and averaging out the obtained scores. 
% While the number of objective quality metrics for point clouds in the state of the art is still limited, a larger number of quality metrics for images have been presented.
% In this context, objective image quality metrics may be adopted to assess the visual quality of point clouds by first creating images from multiple point cloud views, and then computing the image objective metrics on such images. 
% Among the most widely used image quality metrics, the Peak Signal to Noise Ratio (PSNR) measure the difference between two images by mean of the Mean Squared Error (MSE), i.e. by considering the pixel-by-pixel difference between two images. 
% Among the most widely used image quality metrics, 
Image-based PSNR takes into account only pixel-by-pixel color differences between two images while the PSNR-HVS-M~\cite{ponomarenko2007between} also considers the characteristics of the Human Visual System (HVS). 
Other predictors are based on the Structural Similarity Index Measure (SSIM)~\cite{wang2004image}, 
% This metric has, in fact, being proved to be better correlated to subjective scores than the PSNR. 
including different variations such as the Multi-Scale Structural Similarity Index (MS-SSIM)~\cite{wang2003multiscale} 
%which computes the SSIM at different resolutions, 
or the Information Content Weighted Structural Similarity Measure (IW-SSIM) \cite{wang2010information}. % which weights the SSIM scores based on the local information contained in the image. 
%The Visual Information Fidelity (VIF)~\cite{sheik2006visual} uses statistics on natural images to estimate the visual quality of images. Notably, this metrics was adopted as part of the video quality metric Video Multi-Method Assessment Fusion (VMAF)~\cite{li2016toward}, which also includes a Detail Loss Metric (DLM) and a Mean Co-Located Pixel Difference (MCPD).
%The Feature-Similarity Index Metric (FSIM)~\cite{zhang2011fsim} adopts the Phase Congruency (PC) and Gradient Magnitude (GM) to estimate the visual quality of the images. Finally, the Normalized Laplacian Pyramid Distance (NLPD)~\cite{laparra2016perceptual} estimate the visual quality by adopting the Laplacian pyramid to decompose the images.
% Moreover, a large number of objective image quality metrics have been proposed, 
Alternative image objective metrics include Visual Information Fidelity (VIF)~\cite{sheik2006visual}, Feature-Similarity Index Metric (FSIM)~\cite{zhang2011fsim}, and Normalized Laplacian Pyramid Distance (NLPD)~\cite{laparra2016perceptual}. Objective video quality metrics such as the Video Multi-Method Assessment Fusion (VMAF)~\cite{li2016toward} may also be used to assess the visual quality of images.

\subsection{Subjective quality assessment}

Numerous studies evaluating the subjective visual quality of point clouds have been described in recent literature. 
The IRPC dataset \cite{javaheri2020point} employed both G-PCC and V-PCC %together with the PCL [CITE] compression method 
to generate the evaluated stimuli, while \cite{zerman2019subjective} used only V-PCC with dynamic point cloud sequences. 
A comprehensive study produced the M-PCCD dataset \cite{alexiou2019comprehensive} including 9 original point clouds compressed with V-PCC and G-PCC at multiple configurations, 
and a later study conducted as a support for the JPEG Pleno activity \cite{perry2022subjective} employed the same codecs. %, evaluating the performance of recent objective quality metrics such as PointSSIM and PCQM. 

With the rise of deep learning methods for the compression of point clouds, subjective experiments started to include distortions generated with such methods in the set of potential artifacts. 
LB-PCCD \cite{lazzarotto2021benchmarking} used two learning-based codecs during the evaluation and compared them to G-PCC and V-PCC, benchmarking a larger set of quality metrics, while SR-PCD \cite{lazzarotto2022impact} compared one learning-based method to G-PCC using both a standard and a light field monitor, with a later study \cite{lazzarotto2022influence} analyzing the impact of the visualization device on objective metric performance. 
Another experiment \cite{prazeres2022quality2} included four learning-based codecs in the evaluation, and the same author \cite{prazeres2023jpeg} also presented an evaluation of all compression methods proposed for the JPEG Pleno Point Cloud Call for Proposals. 
Recently, MJ-PCCD \cite{lazzarotto2024subjective} compared the performance of G-PCC, V-PCC, and JPEG Pleno, employing three different trade-off strategies between color and geometry compression. %, with a brief discussion on the performance of a small number of quality metrics, but lacking a comprehensive evaluation. 

% Since deep learning has also been used for objective quality evaluation in recent works, datasets with a large number of stimuli with associated subjective scores had to be produced to support neural network training. 
Other datasets including a large number of stimuli with associated subjective scores were produced to support training of learning-based quality metrics. 
The SJTU-PCQA \cite{yang2020predicting} was the result of a subjective experiment conducted to support the development of a learning-based metric, including seven types of distortion but without any of the above-mentioned compression standards. % an using the obtained scores to benchmark a set of image-based metrics. 
The SIAT-PCQD \cite{wu2021subjective} was later produced with stimuli compressed with V-PCC using multiple configurations, introducing parameter values not suggested in the V-PCC Common Test Conditions (CTC) \cite{MPEG-VPCC-CTC} and reference software. %in order to separately evaluate the impact of geometry and color compression on the performance of quality metrics. 
The WPC \cite{liu2022perceptual} dataset included both G-PCC and V-PCC compression artifacts and other distortion types such as downsampling and Gaussian noise. %, evaluating many metrics separating the dataset both according to distortion and content. 
Another large-scale study produced subjective scores for 1240 stimuli obtained from 104 original models in the LS-PCQA dataset \cite{liu2020point}, while the BASICS dataset \cite{ak2023basics} contained 1200 stimuli produced from 75 original point clouds using G-PCC, V-PCC, and learning-based compression.

\begin{table*}[t]
\caption{Performance indexes computed over the entire dataset and after splitting per codec. The color of each cell is associated with its value, where light colors represent high performance and dark colors represent low performance.
%Light colored-cells represent high performance while dark colored-cells represent low performance.
The best-performing metric in each column is highlighted in \textbf{bold}, while the second best is \underline{underlined}.}
\centering
\begin{tabular}{l cccc c cc c cc c cc}
\toprule
& \multicolumn{4}{c}{\textbf{Entire dataset}} && \multicolumn{2}{c}{\textbf{G-PCC}} && \multicolumn{2}{c}{\textbf{V-PCC}} && \multicolumn{2}{c}{\textbf{JPEG Pleno}} \\
  \cmidrule{2-5} \cmidrule{7-8} \cmidrule{10-11} \cmidrule{13-14}
Objective metrics & PLCC & SROCC & RMSE & OR && PLCC & SROCC && PLCC & SROCC && PLCC & SROCC \\
\midrule[0.15em]
D1 PSNR & \cellcolor[HTML]{67CC5C}\textcolor{black}{0.769} & \cellcolor[HTML]{3DBB74}\textcolor{black}{0.686} & \cellcolor[HTML]{F89540}\textcolor{black}{0.749} & \cellcolor[HTML]{C13A50}\textcolor{lightgray}{0.742} && \cellcolor[HTML]{B7DD29}\textcolor{black}{0.891} & \cellcolor[HTML]{90D643}\textcolor{black}{0.835} && \cellcolor[HTML]{64CB5D}\textcolor{black}{0.765} & \cellcolor[HTML]{42BE71}\textcolor{black}{0.696} && \cellcolor[HTML]{7CD24F}\textcolor{black}{0.804} & \cellcolor[HTML]{64CB5D}\textcolor{black}{0.762} \\
D2 PSNR & \cellcolor[HTML]{7ED24E}\textcolor{black}{0.808} & \cellcolor[HTML]{4BC26C}\textcolor{black}{0.717} & \cellcolor[HTML]{FCA934}\textcolor{black}{0.691} & \cellcolor[HTML]{DA4E3B}\textcolor{black}{0.704} && \cellcolor[HTML]{CFE11C}\textcolor{black}{0.927} & \cellcolor[HTML]{A7DB33}\textcolor{black}{0.868} && \cellcolor[HTML]{59C764}\textcolor{black}{0.745} & \cellcolor[HTML]{40BD72}\textcolor{black}{0.692} && \cellcolor[HTML]{95D73F}\textcolor{black}{0.842} & \cellcolor[HTML]{72CF55}\textcolor{black}{0.787} \\
D3 PSNR & \cellcolor[HTML]{30B47A}\textcolor{black}{0.655} & \cellcolor[HTML]{345F8D}\textcolor{lightgray}{0.303} & \cellcolor[HTML]{E3685E}\textcolor{black}{0.886} & \cellcolor[HTML]{85206A}\textcolor{lightgray}{0.817} && \cellcolor[HTML]{92D741}\textcolor{black}{0.840} & \cellcolor[HTML]{90D643}\textcolor{black}{0.835} && \cellcolor[HTML]{1E9B89}\textcolor{black}{0.549} & \cellcolor[HTML]{26AC81}\textcolor{black}{0.617} && \cellcolor[HTML]{59C764}\textcolor{black}{0.744} & \cellcolor[HTML]{5EC961}\textcolor{black}{0.753} \\
LogP2D-G & \cellcolor[HTML]{38B976}\textcolor{black}{0.673} & \cellcolor[HTML]{20A485}\textcolor{black}{0.584} & \cellcolor[HTML]{E76E5A}\textcolor{black}{0.867} & \cellcolor[HTML]{AB2E5D}\textcolor{lightgray}{0.770} && \cellcolor[HTML]{70CE56}\textcolor{black}{0.783} & \cellcolor[HTML]{62CA5F}\textcolor{black}{0.762} && \cellcolor[HTML]{453580}\textcolor{lightgray}{0.156} & \cellcolor[HTML]{460B5E}\textcolor{lightgray}{0.031} && \cellcolor[HTML]{90D643}\textcolor{black}{0.835} & \cellcolor[HTML]{97D83E}\textcolor{black}{0.846} \\
MPED & \cellcolor[HTML]{81D34C}\textcolor{black}{0.811} & \cellcolor[HTML]{55C666}\textcolor{black}{0.735} & \cellcolor[HTML]{FCAC32}\textcolor{black}{0.687} & \cellcolor[HTML]{F0701E}\textcolor{black}{0.657} && \cellcolor[HTML]{CDE01D}\textcolor{black}{0.924} & \cellcolor[HTML]{AFDC2E}\textcolor{black}{0.880} && \cellcolor[HTML]{47C06E}\textcolor{black}{0.708} & \cellcolor[HTML]{25AB81}\textcolor{black}{0.616} && \cellcolor[HTML]{97D83E}\textcolor{black}{0.846} & \cellcolor[HTML]{83D34B}\textcolor{black}{0.815} \\
\hdashline[1pt/2pt]
Y PSNR & \cellcolor[HTML]{26AC81}\textcolor{black}{0.619} & \cellcolor[HTML]{1E9A89}\textcolor{black}{0.545} & \cellcolor[HTML]{DC5E66}\textcolor{black}{0.921} & \cellcolor[HTML]{902468}\textcolor{lightgray}{0.803} && \cellcolor[HTML]{2EB27C}\textcolor{black}{0.648} & \cellcolor[HTML]{1FA286}\textcolor{black}{0.577} && \cellcolor[HTML]{62CA5F}\textcolor{black}{0.760} & \cellcolor[HTML]{4BC26C}\textcolor{black}{0.718} && \cellcolor[HTML]{1F938B}\textcolor{black}{0.513} & \cellcolor[HTML]{277C8E}\textcolor{lightgray}{0.419} \\
YUV PSNR & \cellcolor[HTML]{20A585}\textcolor{black}{0.587} & \cellcolor[HTML]{1F928C}\textcolor{black}{0.509} & \cellcolor[HTML]{D6556D}\textcolor{black}{0.950} & \cellcolor[HTML]{952666}\textcolor{lightgray}{0.798} && \cellcolor[HTML]{29AF7F}\textcolor{black}{0.630} & \cellcolor[HTML]{1E998A}\textcolor{black}{0.542} && \cellcolor[HTML]{51C468}\textcolor{black}{0.730} & \cellcolor[HTML]{44BE70}\textcolor{black}{0.701} && \cellcolor[HTML]{22898D}\textcolor{black}{0.476} & \cellcolor[HTML]{2E6D8E}\textcolor{lightgray}{0.359} \\
Histogram Y & \cellcolor[HTML]{1E988A}\textcolor{black}{0.533} & \cellcolor[HTML]{23878D}\textcolor{black}{0.464} & \cellcolor[HTML]{CC4876}\textcolor{black}{0.993} & \cellcolor[HTML]{D44841}\textcolor{black}{0.714} && \cellcolor[HTML]{20A485}\textcolor{black}{0.585} & \cellcolor[HTML]{1F948B}\textcolor{black}{0.519} && \cellcolor[HTML]{30678D}\textcolor{lightgray}{0.332} & \cellcolor[HTML]{3A538B}\textcolor{lightgray}{0.255} && \cellcolor[HTML]{24AA82}\textcolor{black}{0.613} & \cellcolor[HTML]{24AA82}\textcolor{black}{0.613} \\
Histogram YUV & \cellcolor[HTML]{1E9B89}\textcolor{black}{0.550} & \cellcolor[HTML]{218C8D}\textcolor{black}{0.488} & \cellcolor[HTML]{D04D73}\textcolor{black}{0.980} & \cellcolor[HTML]{DF5436}\textcolor{black}{0.695} && \cellcolor[HTML]{33B679}\textcolor{black}{0.661} & \cellcolor[HTML]{28AE7F}\textcolor{black}{0.627} && \cellcolor[HTML]{37598C}\textcolor{lightgray}{0.280} & \cellcolor[HTML]{440255}\textcolor{lightgray}{0.007} && \cellcolor[HTML]{67CC5C}\textcolor{black}{0.767} & \cellcolor[HTML]{62CA5F}\textcolor{black}{0.759} \\
GraphSIM & \cellcolor[HTML]{97D83E}\textcolor{black}{0.844} & \cellcolor[HTML]{62CA5F}\textcolor{black}{0.758} & \cellcolor[HTML]{FDC328}\textcolor{black}{0.628} & \cellcolor[HTML]{E55B30}\textcolor{black}{0.685} && \cellcolor[HTML]{DFE318}\textcolor{black}{\textbf{0.951}} & \cellcolor[HTML]{C2DF22}\textcolor{black}{\textbf{0.907}} && \cellcolor[HTML]{37588C}\textcolor{lightgray}{0.277} & \cellcolor[HTML]{48196B}\textcolor{lightgray}{0.070} && \cellcolor[HTML]{B7DD29}\textcolor{black}{\textbf{0.891}} & \cellcolor[HTML]{AADB32}\textcolor{black}{\underline{0.874}} \\
MS-GraphSIM & \cellcolor[HTML]{83D34B}\textcolor{black}{0.813} & \cellcolor[HTML]{59C764}\textcolor{black}{0.744} & \cellcolor[HTML]{FCAD31}\textcolor{black}{0.682} & \cellcolor[HTML]{CD4247}\textcolor{black}{0.723} && \cellcolor[HTML]{D2E11B}\textcolor{black}{0.931} & \cellcolor[HTML]{B2DD2C}\textcolor{black}{\underline{0.883}} && \cellcolor[HTML]{30688D}\textcolor{lightgray}{0.336} & \cellcolor[HTML]{462F7C}\textcolor{lightgray}{0.136} && \cellcolor[HTML]{AADB32}\textcolor{black}{0.872} & \cellcolor[HTML]{9FD938}\textcolor{black}{0.859} \\
PointSSIM & \cellcolor[HTML]{21A784}\textcolor{black}{0.597} & \cellcolor[HTML]{23888D}\textcolor{black}{0.467} & \cellcolor[HTML]{D8586A}\textcolor{black}{0.941} & \cellcolor[HTML]{DC5039}\textcolor{black}{0.700} && \cellcolor[HTML]{A7DB33}\textcolor{black}{0.869} & \cellcolor[HTML]{90D643}\textcolor{black}{0.832} && \cellcolor[HTML]{30688D}\textcolor{lightgray}{0.337} & \cellcolor[HTML]{45327F}\textcolor{lightgray}{0.145} && \cellcolor[HTML]{59C764}\textcolor{black}{0.744} & \cellcolor[HTML]{3BBA75}\textcolor{black}{0.681} \\
MS-PointSSIM & \cellcolor[HTML]{9FD938}\textcolor{black}{0.858} & \cellcolor[HTML]{67CC5C}\textcolor{black}{0.769} & \cellcolor[HTML]{FCCC25}\textcolor{black}{0.603} & \cellcolor[HTML]{FBA80D}\textcolor{black}{\textbf{0.596}} && \cellcolor[HTML]{BDDE26}\textcolor{black}{0.899} & \cellcolor[HTML]{8BD546}\textcolor{black}{0.825} && \cellcolor[HTML]{7CD24F}\textcolor{black}{0.804} & \cellcolor[HTML]{42BE71}\textcolor{black}{0.696} && \cellcolor[HTML]{9AD83C}\textcolor{black}{0.852} & \cellcolor[HTML]{7CD24F}\textcolor{black}{0.801} \\
\hdashline[1pt/2pt]
LogP2D-JGY & \cellcolor[HTML]{4DC26B}\textcolor{black}{0.720} & \cellcolor[HTML]{39B976}\textcolor{black}{0.678} & \cellcolor[HTML]{EF7E4E}\textcolor{black}{0.813} & \cellcolor[HTML]{F7850E}\textcolor{black}{0.634} && \cellcolor[HTML]{79D151}\textcolor{black}{0.799} & \cellcolor[HTML]{62CA5F}\textcolor{black}{0.761} && \cellcolor[HTML]{355D8C}\textcolor{lightgray}{0.293} & \cellcolor[HTML]{462F7C}\textcolor{lightgray}{0.135} && \cellcolor[HTML]{B2DD2C}\textcolor{black}{\underline{0.886}} & \cellcolor[HTML]{ADDC30}\textcolor{black}{\textbf{0.876}} \\
PCQM & \cellcolor[HTML]{9FD938}\textcolor{black}{0.858} & \cellcolor[HTML]{6DCE58}\textcolor{black}{\underline{0.779}} & \cellcolor[HTML]{FCCC25}\textcolor{black}{0.602} & \cellcolor[HTML]{F8880C}\textcolor{black}{0.629} && \cellcolor[HTML]{C5DF21}\textcolor{black}{0.913} & \cellcolor[HTML]{92D741}\textcolor{black}{0.838} && \cellcolor[HTML]{9FD938}\textcolor{black}{\textbf{0.859}} & \cellcolor[HTML]{5EC961}\textcolor{black}{\underline{0.752}} && \cellcolor[HTML]{8BD546}\textcolor{black}{0.827} & \cellcolor[HTML]{69CC5B}\textcolor{black}{0.773} \\
\hdashline[1pt/2pt]
PCM\_RR & \cellcolor[HTML]{2AB07E}\textcolor{black}{0.636} & \cellcolor[HTML]{208F8C}\textcolor{black}{0.497} & \cellcolor[HTML]{DF6262}\textcolor{black}{0.905} & \cellcolor[HTML]{BA3655}\textcolor{lightgray}{0.751} && \cellcolor[HTML]{2CB17D}\textcolor{black}{0.641} & \cellcolor[HTML]{1E9D88}\textcolor{black}{0.556} && \cellcolor[HTML]{2A778E}\textcolor{lightgray}{0.396} & \cellcolor[HTML]{3F4788}\textcolor{lightgray}{0.211} && \cellcolor[HTML]{57C665}\textcolor{black}{0.739} & \cellcolor[HTML]{4DC26B}\textcolor{black}{0.721} \\
RR-CAP & \cellcolor[HTML]{53C567}\textcolor{black}{0.732} & \cellcolor[HTML]{1E9B89}\textcolor{black}{0.551} & \cellcolor[HTML]{F2844B}\textcolor{black}{0.800} & \cellcolor[HTML]{C43C4E}\textcolor{black}{0.737} && \cellcolor[HTML]{B5DD2B}\textcolor{black}{0.887} & \cellcolor[HTML]{45BF6F}\textcolor{black}{0.704} && \cellcolor[HTML]{1E998A}\textcolor{black}{0.539} & \cellcolor[HTML]{31648D}\textcolor{lightgray}{0.320} && \cellcolor[HTML]{27AD80}\textcolor{black}{0.624} & \cellcolor[HTML]{20918C}\textcolor{black}{0.506} \\
\hdashline[1pt/2pt]
MM-PCQA & \cellcolor[HTML]{47C06E}\textcolor{black}{0.710} & \cellcolor[HTML]{23A982}\textcolor{black}{0.607} & \cellcolor[HTML]{ED7B51}\textcolor{black}{0.825} & \cellcolor[HTML]{AB2E5D}\textcolor{lightgray}{0.770} && \cellcolor[HTML]{AADB32}\textcolor{black}{0.872} & \cellcolor[HTML]{77D052}\textcolor{black}{0.796} && \cellcolor[HTML]{74D054}\textcolor{black}{0.791} & \cellcolor[HTML]{5BC862}\textcolor{black}{0.746} && \cellcolor[HTML]{6DCE58}\textcolor{black}{0.780} & \cellcolor[HTML]{30B47A}\textcolor{black}{0.654} \\
VQA-PC & \cellcolor[HTML]{20A485}\textcolor{black}{0.583} & \cellcolor[HTML]{1F968B}\textcolor{black}{0.526} & \cellcolor[HTML]{D6556D}\textcolor{black}{0.953} & \cellcolor[HTML]{9C2963}\textcolor{lightgray}{0.789} && \cellcolor[HTML]{92D741}\textcolor{black}{0.839} & \cellcolor[HTML]{6BCD59}\textcolor{black}{0.777} && \cellcolor[HTML]{26808E}\textcolor{black}{0.435} & \cellcolor[HTML]{2D6E8E}\textcolor{lightgray}{0.360} && \cellcolor[HTML]{23A982}\textcolor{black}{0.608} & \cellcolor[HTML]{1F938B}\textcolor{black}{0.514} \\
\hdashline[1pt/2pt]
Image Y PSNR & \cellcolor[HTML]{25AB81}\textcolor{black}{0.616} & \cellcolor[HTML]{1E9C89}\textcolor{black}{0.554} & \cellcolor[HTML]{DC5D66}\textcolor{black}{0.924} & \cellcolor[HTML]{892269}\textcolor{lightgray}{0.812} && \cellcolor[HTML]{4DC26B}\textcolor{black}{0.721} & \cellcolor[HTML]{2CB17D}\textcolor{black}{0.643} && \cellcolor[HTML]{1E9D88}\textcolor{black}{0.558} & \cellcolor[HTML]{1E9A89}\textcolor{black}{0.545} && \cellcolor[HTML]{1E9C89}\textcolor{black}{0.554} & \cellcolor[HTML]{1E998A}\textcolor{black}{0.541} \\
Image YUV PSNR & \cellcolor[HTML]{3BBA75}\textcolor{black}{0.683} & \cellcolor[HTML]{1FA386}\textcolor{black}{0.581} & \cellcolor[HTML]{E87158}\textcolor{black}{0.857} & \cellcolor[HTML]{A32B61}\textcolor{lightgray}{0.779} && \cellcolor[HTML]{5EC961}\textcolor{black}{0.753} & \cellcolor[HTML]{32B57A}\textcolor{black}{0.657} && \cellcolor[HTML]{39B976}\textcolor{black}{0.677} & \cellcolor[HTML]{1E9F88}\textcolor{black}{0.565} && \cellcolor[HTML]{23A883}\textcolor{black}{0.602} & \cellcolor[HTML]{1E978A}\textcolor{black}{0.529} \\
Image MS-SSIM & \cellcolor[HTML]{95D73F}\textcolor{black}{0.842} & \cellcolor[HTML]{57C665}\textcolor{black}{0.739} & \cellcolor[HTML]{FDC128}\textcolor{black}{0.632} & \cellcolor[HTML]{F37918}\textcolor{black}{0.648} && \cellcolor[HTML]{D2E11B}\textcolor{black}{0.933} & \cellcolor[HTML]{86D449}\textcolor{black}{0.820} && \cellcolor[HTML]{40BD72}\textcolor{black}{0.694} & \cellcolor[HTML]{27AD80}\textcolor{black}{0.624} && \cellcolor[HTML]{88D547}\textcolor{black}{0.822} & \cellcolor[HTML]{57C665}\textcolor{black}{0.740} \\
Image IW-SSIM & \cellcolor[HTML]{9AD83C}\textcolor{black}{0.848} & \cellcolor[HTML]{5BC862}\textcolor{black}{0.748} & \cellcolor[HTML]{FDC427}\textcolor{black}{0.621} & \cellcolor[HTML]{F7850E}\textcolor{black}{0.634} && \cellcolor[HTML]{D4E11A}\textcolor{black}{0.937} & \cellcolor[HTML]{8BD546}\textcolor{black}{0.825} && \cellcolor[HTML]{44BE70}\textcolor{black}{0.700} & \cellcolor[HTML]{28AE7F}\textcolor{black}{0.627} && \cellcolor[HTML]{92D741}\textcolor{black}{0.837} & \cellcolor[HTML]{5EC961}\textcolor{black}{0.752} \\
Image VMAF & \cellcolor[HTML]{A2DA37}\textcolor{black}{\underline{0.862}} & \cellcolor[HTML]{69CC5B}\textcolor{black}{0.770} & \cellcolor[HTML]{FBD024}\textcolor{black}{\underline{0.595}} & \cellcolor[HTML]{FB9B06}\textcolor{black}{\underline{0.610}} && \cellcolor[HTML]{DFE318}\textcolor{black}{\underline{0.950}} & \cellcolor[HTML]{A2DA37}\textcolor{black}{0.861} && \cellcolor[HTML]{3DBB74}\textcolor{black}{0.684} & \cellcolor[HTML]{23A982}\textcolor{black}{0.607} && \cellcolor[HTML]{92D741}\textcolor{black}{0.837} & \cellcolor[HTML]{70CE56}\textcolor{black}{0.782} \\
Image VIF & \cellcolor[HTML]{79D151}\textcolor{black}{0.797} & \cellcolor[HTML]{38B976}\textcolor{black}{0.675} & \cellcolor[HTML]{FBA337}\textcolor{black}{0.708} & \cellcolor[HTML]{DA4E3B}\textcolor{black}{0.704} && \cellcolor[HTML]{BFDF24}\textcolor{black}{0.904} & \cellcolor[HTML]{77D052}\textcolor{black}{0.794} && \cellcolor[HTML]{32B57A}\textcolor{black}{0.660} & \cellcolor[HTML]{20A585}\textcolor{black}{0.589} && \cellcolor[HTML]{53C567}\textcolor{black}{0.732} & \cellcolor[HTML]{36B877}\textcolor{black}{0.668} \\
Image FSIM & \cellcolor[HTML]{ADDC30}\textcolor{black}{\textbf{0.876}} & \cellcolor[HTML]{74D054}\textcolor{black}{\textbf{0.790}} & \cellcolor[HTML]{F8DD24}\textcolor{black}{\textbf{0.566}} & \cellcolor[HTML]{FA9706}\textcolor{black}{0.615} && \cellcolor[HTML]{D2E11B}\textcolor{black}{0.930} & \cellcolor[HTML]{90D643}\textcolor{black}{0.833} && \cellcolor[HTML]{97D83E}\textcolor{black}{\underline{0.845}} & \cellcolor[HTML]{72CF55}\textcolor{black}{\textbf{0.785}} && \cellcolor[HTML]{A5DA35}\textcolor{black}{0.867} & \cellcolor[HTML]{6DCE58}\textcolor{black}{0.781} \\
Image NLPD & \cellcolor[HTML]{67CC5C}\textcolor{black}{0.769} & \cellcolor[HTML]{36B877}\textcolor{black}{0.669} & \cellcolor[HTML]{F89540}\textcolor{black}{0.750} & \cellcolor[HTML]{C83E4B}\textcolor{black}{0.732} && \cellcolor[HTML]{B2DD2C}\textcolor{black}{0.884} & \cellcolor[HTML]{70CE56}\textcolor{black}{0.782} && \cellcolor[HTML]{2FB37B}\textcolor{black}{0.649} & \cellcolor[HTML]{20A485}\textcolor{black}{0.584} && \cellcolor[HTML]{64CB5D}\textcolor{black}{0.762} & \cellcolor[HTML]{59C764}\textcolor{black}{0.743} \\
Image PSNR HVS & \cellcolor[HTML]{4BC26C}\textcolor{black}{0.718} & \cellcolor[HTML]{2FB37B}\textcolor{black}{0.649} & \cellcolor[HTML]{EF7D4F}\textcolor{black}{0.816} & \cellcolor[HTML]{A72D5F}\textcolor{lightgray}{0.775} && \cellcolor[HTML]{90D643}\textcolor{black}{0.835} & \cellcolor[HTML]{5EC961}\textcolor{black}{0.752} && \cellcolor[HTML]{23A883}\textcolor{black}{0.604} & \cellcolor[HTML]{1FA286}\textcolor{black}{0.576} && \cellcolor[HTML]{4DC26B}\textcolor{black}{0.720} & \cellcolor[HTML]{2EB27C}\textcolor{black}{0.645} \\\bottomrule
\end{tabular}
\label{tab:entire_dataset_correlation}
\end{table*}

\section{Data collection and processing}
\subsection{Subjective quality experiment}

The MJ-PCCD dataset is used in this study for assessing the performance of objective quality metrics, created by compressing six point clouds from the JPEG Pleno test set \cite{JPEG-Pleno-PC-CTTC} at four different bitrates 
with the JPEG Pleno, G-PCC, and V-PCC standards. 
% The original models can be visualized in Figure \ref{fig:dataset}. 
The set of uncompressed models is composed of four solid point clouds, where points are very closely packed together, and two models with sparser distribution, i.e. \emph{Boxer} and \emph{House\_without\_roof}. 
% The CTC of each  and default configuration parameters were used to define four rates for each codec. 
For G-PCC, the rates $r02$ to $r05$ defined in the CTC \cite{MPEG-GPCC-CTC} were employed, while for V-PCC, the parameters defined for $r1$ to $r4$ were used. 
Following the JPEG Pleno Common Training and Test Conditions \cite{JPEG-Pleno-PC-CTTC}, the configuration parameters were tuned separately for each point cloud to match the bitrate of G-PCC. 
Moreover, three rate allocation strategies P1, P2, and P3 were defined. 
For both G-PCC and V-PCC, the configuration parameters from the CTC were set as P1. 
For G-PCC, both P2 and P3 assigned a higher weight of the bitrate to the color representation while keeping the same bitrate as P1. 
% For V-PCC, it was found that it was not practical to increase the bitrate of the color representation at the same rate given how it would impact the geometry quality. 
For V-PCC, P2 was defined to assign a higher portion of the bitrate to geometry, while P3 was set to use a higher resolution for the occupancy map while adapting the geometry quality parameter accordingly, in both cases keeping the same bitrate as P1. 
For JPEG Pleno, since no standard rate allocation method is defined, the three rate allocation strategies were defined according to the percentage of the bitrate assigned to the geometry representation, with less than 40\% being used in P1, more than 60\% being used in P3, and an intermediary value for P2. 
The entire process produced a total of 213 distorted stimuli. %, with more details on the process being available in [CITE]. 
The rationale behind the choices for P1, P2, and P3 is thoroughly discussed in \cite{lazzarotto2024subjective}.

Subjective visual quality scores were collected during a viewing session with 20 naive subjects in a controlled lab environment. The simultaneous DSIS methodology with hidden reference was adopted for the experiment, by employing the Unity framework presented in~\cite{lazzarotto2022impact}. The experiment was conducted in a \emph{passive} manner, where the test subjects observed the point cloud pairs simultaneously rotating 360\textdegree around their vertical axis for 12 seconds. Afterward, the interface would move to the voting interface, where subjects were asked to rate the impairment between the two point clouds on a 5-level discrete quality scale. The point clouds were displayed in a DELL UltraSharp U3219Q monitor with 31.5 inches of diagonal size, calibrated using the guidelines provided in ITU-R Recommendations BT.500~\cite{bt.500}. The room lighting conditions were set to a value of approximately 15 lux, while the viewing distance was set to approximately 3.2H \cite{bt.500}, corresponding to 48 cm. 
The collected subjective scores were first reviewed for outliers following the procedure in BT.500~\cite{bt.500}, which did not reveal any outliers. All subjective raw scores were then processed by computing the MOS.

\subsection{Objective quality metrics}

A large set of objective quality metrics were computed on the distorted point clouds. % evaluated in the subjective experiment. 
Point-to-point PSNR (D1 PSNR) and point-to-plane PSNR \cite{tian2017geometric} (D2 PSNR), as well as point-based color PSNR, were computed using the software provided by MPEG. 
For the latter, both the value computed only on the luminance channel (Y PSNR) and a weighted average between luminance and chrominance PSNR (YUV PSNR) were retained, the latter using a weighted average where Y, U, and V received a relative weight of 6, 1, and 1 respectively. 
The PSNR value for the density-to-density (D3 PSNR) \cite{m61195} was also computed with software made available in the MPEG repository. 
Both the geometry-only \cite{javaheri2020point} and joint logarithmic \cite{javaheri2021point} point-to-distribution metrics were computed with source code made available by the authors, being referred to as LogP2D-G and LogP2D-JGY, respectively. 
The color-based histogram metric \cite{viola2020color} was also computed, using L2 as the distance estimator both for the Y and YUV channels, the latter with the same weights as YUV PSNR, being referred to as Histogram Y and Histogram YUV. 
% PointSSIM was also computed using default parameters, i.e. using k=12 to form local neighbors, luminance as an attribute, variance as an estimator, and the mean as the pooling method, without voxelization pre-processing. 
The source code provided by the authors and default configurations were used for the remaining metrics, i.e. MPED, GraphSIM, MS-GraphSIM, PointSSIM, MS-PointSSIM, PCQM, PCM\_RR, RR-CAP, MM-PCQA and VQA-PC.

In addition to the above point cloud quality metrics, this study also included image quality metrics computed in projections of the point clouds at different angles. 
The framework used to conduct the subjective experiment was used to render the projections, keeping the same point size as the experiment. 
Since the point clouds were rotated along their vertical axis during the inspection, four projections were rendered for each stimulus at the rotation angles of 0\textdegree, 90\textdegree, 180\textdegree, and 270\textdegree. 
Each projection was then cropped to remove most of the background, with the same cropping area being kept across all the stimuli generated from the same original point cloud for a given rotation angle. 
The following FR image-based metrics were then computed using the JPEG AI Quality Assessment Framework %\footnote{\url{https://gitlab.com/wg1/jpeg-ai/jpeg-ai-qaf}} 
between the projections of the distorted stimuli and the reference: Y PSNR, YUV PSNR, MS-SSIM, IW-SSIM, VMAF, VIF, FSIM, NLPD and PSNR HVS. 
% As a result, each stimulus received four scores per metric, one for each projection angle
The average between the scores across the four projection angles was employed as a final quality predictor for each metric. 
For the metrics designed for the YUV colorspace, the convention provided in ITU-R Recommendations BT.709~\cite{BT709} was used for conversion from RGB. 
% It is worth noting that the MS-SSIM, IW-SSIM, PSNR-HVS-M, VIF, and NLPD were designed for singe-channel images. On the other hand, VMAF takes as input images in the YUV color space. Therefore color conversion were applied, where the RGB images were converted to the luminance channel Y or to the YUV color space by using the convention provided in ITU-R Recommendations BT.709~\cite{BT709}.

\subsection{Benchmarking settings}

Prior to the assessment of the quality predictors, the objective scores were fitted to the MOS following the procedure reported in~\cite{ITUTutorial} to generate a predicted MOS value (PMOS), adopting the logistic function with offset depicted in Equation \ref{eq:logistic}. 

\begin{equation}
    PMOS=S_p=\beta_1+\frac{\beta_2}{1+\exp(-\beta_3*(S-\beta_4))}
    \label{eq:logistic}
\end{equation}

The performance of the objective quality metrics was then evaluated through multiple performance indexes, including the correlation between MOS and PMOS using both the Pearson Linear Correlation Coefficient (PLCC) and the Spearman's Rank Correlation Coefficient (SROCC). 
In addition, the Root Mean Square Error (RMSE) and the Outlier Ratio (OR) were also computed following the guidelines in ITU-T Recommendation P.1401~\cite{P1401}.
% have been computed to determine the correlation between the two variables. 
% As the sign of the correlation coefficient is not relevant in this analysis, only absolute values were considered. 

% It is well-known that MOS scores tend to be non-linear at the extremes of the quality scale, causing the relationship between the MOS and objective scores to be non-linear. 
% As the PLCC assumes linear correlation, a least-squares regression procedure was applied to the data in order to remove this non-linearity, by following the procedure reported in the ITU Tutorial ``Objective Perceptual Assessment of Video Quality"~\cite{ITUTutorial}. For this purpose, a 4-parameter logistic curve with offset was adopted:
% \begin{equation}
%     S_p=\beta_1+\frac{\beta_2}{1+\exp(-\beta_3*(S-\beta_4))}
% \end{equation}

% Add that the correlation was computed separating the dataset

While the performance indexes computed over the entire dataset illustrate the overall performance of each metric, a more detailed analysis of their correlation with subjective scores in defined conditions can provide additional insights. 
% a better idea of their performance in specific situations. 
For this reason, the PLCC and SROCC are also computed after splitting the dataset according to the codec and to the content, after separately computing the fitting function in each subset. 

\begin{figure*}
    \centering

        \includegraphics[width=0.19\linewidth]{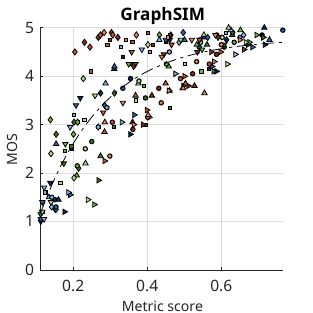}
        \includegraphics[width=0.19\linewidth]{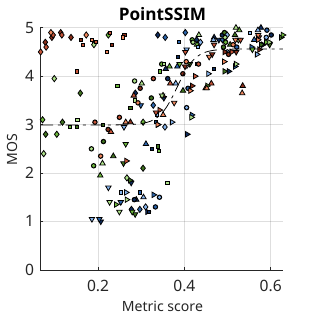}
        \includegraphics[width=0.19\linewidth]{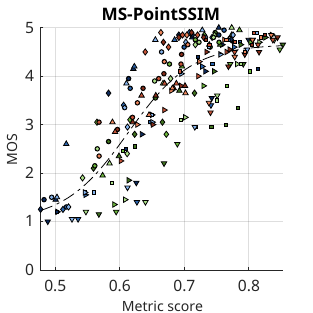}
        \includegraphics[width=0.19\linewidth]{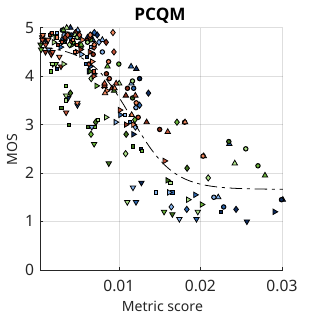}
        \begin{minipage}[b]{0.19\linewidth}
        \centering
            \includegraphics[width=0.5\linewidth]{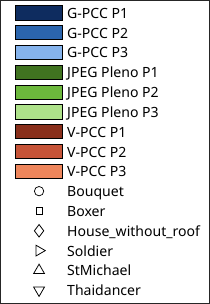}
            \vspace{17px}
        \end{minipage}
        
    \caption{Scatter plots of objective against subjective scores for different quality metrics, with the logistic fitting curve}
    \label{fig:scatter_plots}
\end{figure*}

\section{Results and discussion}

The obtained performance indexes for the entire dataset are displayed in the first four columns of Table \ref{tab:entire_dataset_correlation}. 
None of the analyzed metrics is able to achieve PLCC higher than 0.9 or SROCC higher than 0.8, suggesting that current metrics are still not able to faithfully model the human visual system in the 3D domain. 
The image-based FSIM metric achieves the best performance index values, followed by VMAF, PCQM, and MS-PointSSIM. 
The advantage of using image quality metrics in this case may be because they take into account the information on how the point clouds were rendered. 
In fact, changing the point size on the subjective assessment platform could have an impact on the subjective scores, which would be reflected in image metrics while point cloud metrics would remain unchanged. 
Apart from FSIM and VMAF, MS-SSIM and IW-SSIM are also able to achieve good performance. 

Regarding point cloud metrics, PCQM and MS-PointSSIM achieve the highest performances, the first with an advantage on SROCC and the latter with lower OR. 
GraphSIM and MS-GraphSIM come next, displaying better performance than the two best geometry-based metrics, i.e. MPED and D2 PSNR. 
However, while the multi-scale paradigm allows MS-PointSSIM to outperform PointSSIM by a large margin, it does not have the same effect on GraphSIM, which achieves a slightly higher performance than MS-GraphSIM. 
Many of the observed results for FR metrics agree with previous studies, such as the superiority of PCQM and GraphSIM against geometry-only metrics \cite{lazzarotto2021benchmarking, lazzarotto2022influence}. 
The low performance of PointSSIM appears to agree with studies conducted with larger datasets \cite{liu2022perceptual}, differing from other evaluations which found its performance to be higher \cite{lazzarotto2021benchmarking, lazzarotto2022influence}. 
As expected, RR and NR quality metrics display lower performance than the evaluated FR metrics. %, mainly due to their lack of information about the reference. 
Within those categories, the use of multi-modal learning allowed MM-PCQA to outperform VQA-PC, which relies only on projections onto a moving camera. 
The use of 2D saliency maps on RR-CAP allowed it to reach higher performance compared to the 3D-based features used by PCM\_RR. % obtained in the 3D domain. 
Figure \ref{fig:scatter_plots} displays the scatter plots of some objective scores against the MOS. %, further illustrating the predictive ability of the best-performing metrics as well as other cases of interest. 

\begin{table*}[t]
\caption{Performance indexes computed after splitting per content. The color of each cell is associated with its value, where light colors represent high performance and dark colors represent low performance. 
%Light colored-cells represent high performance while dark colored-cells represent low performance.
The best-performing metric in each column is highlighted in \textbf{bold}, while the second best is \underline{underlined}.}
\centering
\resizebox{\textwidth}{!}{
\begin{tabular}{l cc c cc c cc c cc c cc c cc}
\toprule
& \multicolumn{2}{c}{\textbf{Bouquet}} && \multicolumn{2}{c}{\textbf{Boxer}} && \multicolumn{2}{c}{\textbf{House\_without\_roof}} && \multicolumn{2}{c}{\textbf{Soldier}} && \multicolumn{2}{c}{\textbf{StMichael}} && \multicolumn{2}{c}{\textbf{Thaidancer}} \\
  \cmidrule{2-3} \cmidrule{5-6} \cmidrule{8-9} \cmidrule{11-12} \cmidrule{14-15} \cmidrule{17-18}
Objective metrics & PLCC & SROCC && PLCC & SROCC && PLCC & SROCC && PLCC & SROCC && PLCC & SROCC && PLCC & SROCC \\
\midrule[0.15em]
D1 PSNR & \cellcolor[HTML]{8BD546}\textcolor{black}{0.824} & \cellcolor[HTML]{64CB5D}\textcolor{black}{0.765} && \cellcolor[HTML]{86D449}\textcolor{black}{0.820} & \cellcolor[HTML]{3EBC73}\textcolor{black}{0.688} && \cellcolor[HTML]{A7DB33}\textcolor{black}{0.868} & \cellcolor[HTML]{3BBA75}\textcolor{black}{0.683} && \cellcolor[HTML]{57C665}\textcolor{black}{0.739} & \cellcolor[HTML]{5EC961}\textcolor{black}{0.752} && \cellcolor[HTML]{40BD72}\textcolor{black}{0.693} & \cellcolor[HTML]{2EB27C}\textcolor{black}{0.645} && \cellcolor[HTML]{CFE11C}\textcolor{black}{0.927} & \cellcolor[HTML]{95D73F}\textcolor{black}{0.842} \\
D2 PSNR & \cellcolor[HTML]{A2DA37}\textcolor{black}{0.862} & \cellcolor[HTML]{86D449}\textcolor{black}{0.818} && \cellcolor[HTML]{8BD546}\textcolor{black}{0.827} & \cellcolor[HTML]{44BE70}\textcolor{black}{0.702} && \cellcolor[HTML]{88D547}\textcolor{black}{0.824} & \cellcolor[HTML]{2FB37B}\textcolor{black}{0.650} && \cellcolor[HTML]{70CE56}\textcolor{black}{0.784} & \cellcolor[HTML]{7ED24E}\textcolor{black}{0.805} && \cellcolor[HTML]{5EC961}\textcolor{black}{0.750} & \cellcolor[HTML]{53C567}\textcolor{black}{0.732} && \cellcolor[HTML]{DFE318}\textcolor{black}{0.950} & \cellcolor[HTML]{9DD93A}\textcolor{black}{0.853} \\
D3 PSNR & \cellcolor[HTML]{21A685}\textcolor{black}{0.591} & \cellcolor[HTML]{20A585}\textcolor{black}{0.588} && \cellcolor[HTML]{7CD24F}\textcolor{black}{0.804} & \cellcolor[HTML]{2E6C8E}\textcolor{lightgray}{0.353} && \cellcolor[HTML]{67CC5C}\textcolor{black}{0.767} & \cellcolor[HTML]{365B8C}\textcolor{lightgray}{0.287} && \cellcolor[HTML]{218C8D}\textcolor{black}{0.487} & \cellcolor[HTML]{1EA087}\textcolor{black}{0.569} && \cellcolor[HTML]{20908C}\textcolor{black}{0.502} & \cellcolor[HTML]{2D6E8E}\textcolor{lightgray}{0.360} && \cellcolor[HTML]{4DC26B}\textcolor{black}{0.722} & \cellcolor[HTML]{60C960}\textcolor{black}{0.756} \\
LogP2D-G & \cellcolor[HTML]{86D449}\textcolor{black}{0.818} & \cellcolor[HTML]{5EC961}\textcolor{black}{0.752} && \cellcolor[HTML]{3BBA75}\textcolor{black}{0.681} & \cellcolor[HTML]{26AC81}\textcolor{black}{0.620} && \cellcolor[HTML]{47C06E}\textcolor{black}{0.711} & \cellcolor[HTML]{1E988A}\textcolor{black}{0.532} && \cellcolor[HTML]{53C567}\textcolor{black}{0.734} & \cellcolor[HTML]{59C764}\textcolor{black}{0.745} && \cellcolor[HTML]{28AE7F}\textcolor{black}{0.626} & \cellcolor[HTML]{1E978A}\textcolor{black}{0.530} && \cellcolor[HTML]{CFE11C}\textcolor{black}{0.929} & \cellcolor[HTML]{9AD83C}\textcolor{black}{0.849} \\
MPED & \cellcolor[HTML]{88D547}\textcolor{black}{0.822} & \cellcolor[HTML]{7CD24F}\textcolor{black}{0.804} && \cellcolor[HTML]{8DD644}\textcolor{black}{0.830} & \cellcolor[HTML]{39B976}\textcolor{black}{0.678} && \cellcolor[HTML]{C7E01F}\textcolor{black}{0.915} & \cellcolor[HTML]{72CF55}\textcolor{black}{0.788} && \cellcolor[HTML]{86D449}\textcolor{black}{0.819} & \cellcolor[HTML]{AADB32}\textcolor{black}{0.874} && \cellcolor[HTML]{40BD72}\textcolor{black}{0.695} & \cellcolor[HTML]{42BE71}\textcolor{black}{0.695} && \cellcolor[HTML]{C2DF22}\textcolor{black}{0.909} & \cellcolor[HTML]{AADB32}\textcolor{black}{0.872} \\
\hdashline[1pt/2pt]
Y PSNR & \cellcolor[HTML]{E1E318}\textcolor{black}{0.954} & \cellcolor[HTML]{BADE27}\textcolor{black}{0.898} && \cellcolor[HTML]{A7DB33}\textcolor{black}{0.869} & \cellcolor[HTML]{39B976}\textcolor{black}{0.676} && \cellcolor[HTML]{BADE27}\textcolor{black}{0.898} & \cellcolor[HTML]{81D34C}\textcolor{black}{0.810} && \cellcolor[HTML]{FAE622}\textcolor{black}{\textbf{0.995}} & \cellcolor[HTML]{F8E621}\textcolor{black}{\textbf{0.989}} && \cellcolor[HTML]{C5DF21}\textcolor{black}{0.911} & \cellcolor[HTML]{95D73F}\textcolor{black}{0.843} && \cellcolor[HTML]{F8E621}\textcolor{black}{0.989} & \cellcolor[HTML]{E1E318}\textcolor{black}{0.956} \\
YUV PSNR & \cellcolor[HTML]{CFE11C}\textcolor{black}{0.926} & \cellcolor[HTML]{B2DD2C}\textcolor{black}{0.884} && \cellcolor[HTML]{AADB32}\textcolor{black}{0.872} & \cellcolor[HTML]{44BE70}\textcolor{black}{0.703} && \cellcolor[HTML]{BADE27}\textcolor{black}{0.897} & \cellcolor[HTML]{83D34B}\textcolor{black}{0.814} && \cellcolor[HTML]{F3E51E}\textcolor{black}{0.983} & \cellcolor[HTML]{ECE41A}\textcolor{black}{0.971} && \cellcolor[HTML]{C5DF21}\textcolor{black}{0.911} & \cellcolor[HTML]{95D73F}\textcolor{black}{0.844} && \cellcolor[HTML]{F6E61F}\textcolor{black}{0.985} & \cellcolor[HTML]{E4E318}\textcolor{black}{0.960} \\
Histogram Y & \cellcolor[HTML]{CFE11C}\textcolor{black}{0.927} & \cellcolor[HTML]{A2DA37}\textcolor{black}{0.863} && \cellcolor[HTML]{8DD644}\textcolor{black}{0.831} & \cellcolor[HTML]{5EC961}\textcolor{black}{0.753} && \cellcolor[HTML]{81D34C}\textcolor{black}{0.809} & \cellcolor[HTML]{24858D}\textcolor{black}{0.457} && \cellcolor[HTML]{ECE41A}\textcolor{black}{0.972} & \cellcolor[HTML]{D4E11A}\textcolor{black}{0.937} && \cellcolor[HTML]{E1E318}\textcolor{black}{\textbf{0.953}} & \cellcolor[HTML]{DFE318}\textcolor{black}{\textbf{0.952}} && \cellcolor[HTML]{F3E51E}\textcolor{black}{0.982} & \cellcolor[HTML]{D7E219}\textcolor{black}{0.941} \\
Histogram YUV & \cellcolor[HTML]{ADDC30}\textcolor{black}{0.876} & \cellcolor[HTML]{BFDF24}\textcolor{black}{0.906} && \cellcolor[HTML]{33608D}\textcolor{lightgray}{0.306} & \cellcolor[HTML]{3E4989}\textcolor{lightgray}{0.222} && \cellcolor[HTML]{62CA5F}\textcolor{black}{0.760} & \cellcolor[HTML]{21A784}\textcolor{black}{0.594} && \cellcolor[HTML]{A2DA37}\textcolor{black}{0.860} & \cellcolor[HTML]{62CA5F}\textcolor{black}{0.759} && \cellcolor[HTML]{ADDC30}\textcolor{black}{0.876} & \cellcolor[HTML]{A7DB33}\textcolor{black}{0.869} && \cellcolor[HTML]{E7E419}\textcolor{black}{0.961} & \cellcolor[HTML]{D2E11B}\textcolor{black}{0.931} \\
GraphSIM & \cellcolor[HTML]{E4E318}\textcolor{black}{0.961} & \cellcolor[HTML]{DCE218}\textcolor{black}{0.946} && \cellcolor[HTML]{ADDC30}\textcolor{black}{0.876} & \cellcolor[HTML]{3EBC73}\textcolor{black}{0.689} && \cellcolor[HTML]{8DD644}\textcolor{black}{0.829} & \cellcolor[HTML]{70CE56}\textcolor{black}{0.782} && \cellcolor[HTML]{F6E61F}\textcolor{black}{0.985} & \cellcolor[HTML]{ECE41A}\textcolor{black}{0.971} && \cellcolor[HTML]{AFDC2E}\textcolor{black}{0.881} & \cellcolor[HTML]{90D643}\textcolor{black}{0.836} && \cellcolor[HTML]{F6E61F}\textcolor{black}{0.987} & \cellcolor[HTML]{E1E318}\textcolor{black}{0.956} \\
MS-GraphSIM & \cellcolor[HTML]{D4E11A}\textcolor{black}{0.935} & \cellcolor[HTML]{CFE11C}\textcolor{black}{0.929} && \cellcolor[HTML]{9DD93A}\textcolor{black}{0.855} & \cellcolor[HTML]{38B976}\textcolor{black}{0.673} && \cellcolor[HTML]{81D34C}\textcolor{black}{0.810} & \cellcolor[HTML]{70CE56}\textcolor{black}{0.783} && \cellcolor[HTML]{E9E419}\textcolor{black}{0.966} & \cellcolor[HTML]{E7E419}\textcolor{black}{0.963} && \cellcolor[HTML]{9AD83C}\textcolor{black}{0.850} & \cellcolor[HTML]{83D34B}\textcolor{black}{0.813} && \cellcolor[HTML]{E4E318}\textcolor{black}{0.960} & \cellcolor[HTML]{C5DF21}\textcolor{black}{0.911} \\
PointSSIM & \cellcolor[HTML]{74D054}\textcolor{black}{0.792} & \cellcolor[HTML]{9FD938}\textcolor{black}{0.855} && \cellcolor[HTML]{472A79}\textcolor{lightgray}{0.118} & \cellcolor[HTML]{471466}\textcolor{lightgray}{0.052} && \cellcolor[HTML]{2F6A8D}\textcolor{lightgray}{0.346} & \cellcolor[HTML]{433A83}\textcolor{lightgray}{0.168} && \cellcolor[HTML]{DFE318}\textcolor{black}{0.952} & \cellcolor[HTML]{ECE41A}\textcolor{black}{\underline{0.971}} && \cellcolor[HTML]{8DD644}\textcolor{black}{0.829} & \cellcolor[HTML]{7CD24F}\textcolor{black}{0.804} && \cellcolor[HTML]{E4E318}\textcolor{black}{0.958} & \cellcolor[HTML]{DAE218}\textcolor{black}{0.943} \\
MS-PointSSIM & \cellcolor[HTML]{F1E51C}\textcolor{black}{\textbf{0.977}} & \cellcolor[HTML]{E9E419}\textcolor{black}{\textbf{0.966}} && \cellcolor[HTML]{B2DD2C}\textcolor{black}{0.886} & \cellcolor[HTML]{70CE56}\textcolor{black}{0.783} && \cellcolor[HTML]{BADE27}\textcolor{black}{0.897} & \cellcolor[HTML]{70CE56}\textcolor{black}{0.784} && \cellcolor[HTML]{ECE41A}\textcolor{black}{0.971} & \cellcolor[HTML]{E7E419}\textcolor{black}{0.961} && \cellcolor[HTML]{BADE27}\textcolor{black}{0.895} & \cellcolor[HTML]{9AD83C}\textcolor{black}{0.850} && \cellcolor[HTML]{FAE622}\textcolor{black}{\textbf{0.993}} & \cellcolor[HTML]{E4E318}\textcolor{black}{\underline{0.960}} \\
\hdashline[1pt/2pt]
LogP2D-JGY & \cellcolor[HTML]{CAE01E}\textcolor{black}{0.920} & \cellcolor[HTML]{DCE218}\textcolor{black}{0.948} && \cellcolor[HTML]{60C960}\textcolor{black}{0.757} & \cellcolor[HTML]{1E9D88}\textcolor{black}{0.558} && \cellcolor[HTML]{23878D}\textcolor{black}{0.461} & \cellcolor[HTML]{287A8E}\textcolor{lightgray}{0.414} && \cellcolor[HTML]{DFE318}\textcolor{black}{0.949} & \cellcolor[HTML]{DAE218}\textcolor{black}{0.945} && \cellcolor[HTML]{C7E01F}\textcolor{black}{0.918} & \cellcolor[HTML]{BDDE26}\textcolor{black}{0.901} && \cellcolor[HTML]{F1E51C}\textcolor{black}{0.980} & \cellcolor[HTML]{DAE218}\textcolor{black}{0.944} \\
PCQM & \cellcolor[HTML]{E4E318}\textcolor{black}{0.958} & \cellcolor[HTML]{E1E318}\textcolor{black}{0.956} && \cellcolor[HTML]{ADDC30}\textcolor{black}{0.876} & \cellcolor[HTML]{62CA5F}\textcolor{black}{0.760} && \cellcolor[HTML]{9DD93A}\textcolor{black}{0.852} & \cellcolor[HTML]{4DC26B}\textcolor{black}{0.720} && \cellcolor[HTML]{F6E61F}\textcolor{black}{\underline{0.986}} & \cellcolor[HTML]{E9E419}\textcolor{black}{0.966} && \cellcolor[HTML]{DFE318}\textcolor{black}{\underline{0.952}} & \cellcolor[HTML]{C5DF21}\textcolor{black}{\underline{0.914}} && \cellcolor[HTML]{FAE622}\textcolor{black}{\underline{0.993}} & \cellcolor[HTML]{E4E318}\textcolor{black}{\textbf{0.960}} \\
\hdashline[1pt/2pt]
PCM\_RR & \cellcolor[HTML]{70CE56}\textcolor{black}{0.784} & \cellcolor[HTML]{39B976}\textcolor{black}{0.676} && \cellcolor[HTML]{B7DD29}\textcolor{black}{0.892} & \cellcolor[HTML]{69CC5B}\textcolor{black}{0.771} && \cellcolor[HTML]{64CB5D}\textcolor{black}{0.764} & \cellcolor[HTML]{3BBA75}\textcolor{black}{0.681} && \cellcolor[HTML]{BADE27}\textcolor{black}{0.897} & \cellcolor[HTML]{9FD938}\textcolor{black}{0.859} && \cellcolor[HTML]{1E9D88}\textcolor{black}{0.556} & \cellcolor[HTML]{1E998A}\textcolor{black}{0.538} && \cellcolor[HTML]{DAE218}\textcolor{black}{0.942} & \cellcolor[HTML]{AFDC2E}\textcolor{black}{0.881} \\
RR-CAP & \cellcolor[HTML]{C7E01F}\textcolor{black}{0.914} & \cellcolor[HTML]{D2E11B}\textcolor{black}{0.930} && \cellcolor[HTML]{CDE01D}\textcolor{black}{0.926} & \cellcolor[HTML]{90D643}\textcolor{black}{0.833} && \cellcolor[HTML]{ADDC30}\textcolor{black}{0.877} & \cellcolor[HTML]{38B976}\textcolor{black}{0.675} && \cellcolor[HTML]{DAE218}\textcolor{black}{0.945} & \cellcolor[HTML]{E1E318}\textcolor{black}{0.955} && \cellcolor[HTML]{90D643}\textcolor{black}{0.834} & \cellcolor[HTML]{45BF6F}\textcolor{black}{0.705} && \cellcolor[HTML]{ECE41A}\textcolor{black}{0.971} & \cellcolor[HTML]{DAE218}\textcolor{black}{0.942} \\
\hdashline[1pt/2pt]
MM-PCQA & \cellcolor[HTML]{47C06E}\textcolor{black}{0.708} & \cellcolor[HTML]{2AB07E}\textcolor{black}{0.636} && \cellcolor[HTML]{A5DA35}\textcolor{black}{0.865} & \cellcolor[HTML]{40BD72}\textcolor{black}{0.694} && \cellcolor[HTML]{86D449}\textcolor{black}{0.819} & \cellcolor[HTML]{25AB81}\textcolor{black}{0.617} && \cellcolor[HTML]{A5DA35}\textcolor{black}{0.866} & \cellcolor[HTML]{B2DD2C}\textcolor{black}{0.883} && \cellcolor[HTML]{1FA386}\textcolor{black}{0.580} & \cellcolor[HTML]{2A778E}\textcolor{lightgray}{0.398} && \cellcolor[HTML]{9AD83C}\textcolor{black}{0.849} & \cellcolor[HTML]{2AB07E}\textcolor{black}{0.636} \\
VQA-PC & \cellcolor[HTML]{DAE218}\textcolor{black}{0.943} & \cellcolor[HTML]{D2E11B}\textcolor{black}{0.932} && \cellcolor[HTML]{27AD80}\textcolor{black}{0.621} & \cellcolor[HTML]{25828E}\textcolor{black}{0.444} && \cellcolor[HTML]{6BCD59}\textcolor{black}{0.774} & \cellcolor[HTML]{3BBA75}\textcolor{black}{0.682} && \cellcolor[HTML]{88D547}\textcolor{black}{0.821} & \cellcolor[HTML]{1FA187}\textcolor{black}{0.572} && \cellcolor[HTML]{AFDC2E}\textcolor{black}{0.880} & \cellcolor[HTML]{9DD93A}\textcolor{black}{0.854} && \cellcolor[HTML]{E1E318}\textcolor{black}{0.955} & \cellcolor[HTML]{4BC26C}\textcolor{black}{0.719} \\
\hdashline[1pt/2pt]
Image Y PSNR & \cellcolor[HTML]{CDE01D}\textcolor{black}{0.924} & \cellcolor[HTML]{C7E01F}\textcolor{black}{0.915} && \cellcolor[HTML]{BADE27}\textcolor{black}{0.898} & \cellcolor[HTML]{9AD83C}\textcolor{black}{0.851} && \cellcolor[HTML]{D7E219}\textcolor{black}{0.938} & \cellcolor[HTML]{9AD83C}\textcolor{black}{0.851} && \cellcolor[HTML]{C7E01F}\textcolor{black}{0.915} & \cellcolor[HTML]{D2E11B}\textcolor{black}{0.932} && \cellcolor[HTML]{ADDC30}\textcolor{black}{0.877} & \cellcolor[HTML]{9DD93A}\textcolor{black}{0.852} && \cellcolor[HTML]{E9E419}\textcolor{black}{0.966} & \cellcolor[HTML]{C2DF22}\textcolor{black}{0.908} \\
Image YUV PSNR & \cellcolor[HTML]{D4E11A}\textcolor{black}{0.935} & \cellcolor[HTML]{D7E219}\textcolor{black}{0.940} && \cellcolor[HTML]{C5DF21}\textcolor{black}{0.914} & \cellcolor[HTML]{A7DB33}\textcolor{black}{0.869} && \cellcolor[HTML]{DCE218}\textcolor{black}{0.946} & \cellcolor[HTML]{A7DB33}\textcolor{black}{0.870} && \cellcolor[HTML]{DCE218}\textcolor{black}{0.947} & \cellcolor[HTML]{E4E318}\textcolor{black}{0.960} && \cellcolor[HTML]{CAE01E}\textcolor{black}{0.921} & \cellcolor[HTML]{B5DD2B}\textcolor{black}{0.889} && \cellcolor[HTML]{EEE51B}\textcolor{black}{0.973} & \cellcolor[HTML]{CDE01D}\textcolor{black}{0.925} \\
Image MS-SSIM & \cellcolor[HTML]{E9E419}\textcolor{black}{0.969} & \cellcolor[HTML]{E7E419}\textcolor{black}{0.961} && \cellcolor[HTML]{DAE218}\textcolor{black}{\textbf{0.945}} & \cellcolor[HTML]{C2DF22}\textcolor{black}{\textbf{0.908}} && \cellcolor[HTML]{D7E219}\textcolor{black}{0.940} & \cellcolor[HTML]{AADB32}\textcolor{black}{\textbf{0.875}} && \cellcolor[HTML]{E7E419}\textcolor{black}{0.961} & \cellcolor[HTML]{E7E419}\textcolor{black}{0.964} && \cellcolor[HTML]{CDE01D}\textcolor{black}{0.922} & \cellcolor[HTML]{AFDC2E}\textcolor{black}{0.880} && \cellcolor[HTML]{EEE51B}\textcolor{black}{0.976} & \cellcolor[HTML]{D7E219}\textcolor{black}{0.940} \\
Image IW-SSIM & \cellcolor[HTML]{ECE41A}\textcolor{black}{\underline{0.969}} & \cellcolor[HTML]{E9E419}\textcolor{black}{\underline{0.965}} && \cellcolor[HTML]{D7E219}\textcolor{black}{\underline{0.940}} & \cellcolor[HTML]{BFDF24}\textcolor{black}{\underline{0.905}} && \cellcolor[HTML]{DAE218}\textcolor{black}{0.943} & \cellcolor[HTML]{A7DB33}\textcolor{black}{0.869} && \cellcolor[HTML]{E4E318}\textcolor{black}{0.958} & \cellcolor[HTML]{E1E318}\textcolor{black}{0.956} && \cellcolor[HTML]{CFE11C}\textcolor{black}{0.927} & \cellcolor[HTML]{B5DD2B}\textcolor{black}{0.889} && \cellcolor[HTML]{F1E51C}\textcolor{black}{0.978} & \cellcolor[HTML]{DAE218}\textcolor{black}{0.945} \\
Image VMAF & \cellcolor[HTML]{E4E318}\textcolor{black}{0.960} & \cellcolor[HTML]{DFE318}\textcolor{black}{0.952} && \cellcolor[HTML]{BDDE26}\textcolor{black}{0.900} & \cellcolor[HTML]{9AD83C}\textcolor{black}{0.851} && \cellcolor[HTML]{D7E219}\textcolor{black}{0.940} & \cellcolor[HTML]{95D73F}\textcolor{black}{0.843} && \cellcolor[HTML]{D4E11A}\textcolor{black}{0.937} & \cellcolor[HTML]{DAE218}\textcolor{black}{0.943} && \cellcolor[HTML]{CFE11C}\textcolor{black}{0.928} & \cellcolor[HTML]{BADE27}\textcolor{black}{0.898} && \cellcolor[HTML]{F3E51E}\textcolor{black}{0.983} & \cellcolor[HTML]{E4E318}\textcolor{black}{0.959} \\
Image VIF & \cellcolor[HTML]{DFE318}\textcolor{black}{0.949} & \cellcolor[HTML]{DCE218}\textcolor{black}{0.948} && \cellcolor[HTML]{C5DF21}\textcolor{black}{0.913} & \cellcolor[HTML]{9DD93A}\textcolor{black}{0.853} && \cellcolor[HTML]{DCE218}\textcolor{black}{\underline{0.947}} & \cellcolor[HTML]{A5DA35}\textcolor{black}{0.864} && \cellcolor[HTML]{D4E11A}\textcolor{black}{0.937} & \cellcolor[HTML]{DCE218}\textcolor{black}{0.947} && \cellcolor[HTML]{A7DB33}\textcolor{black}{0.871} & \cellcolor[HTML]{92D741}\textcolor{black}{0.837} && \cellcolor[HTML]{EEE51B}\textcolor{black}{0.974} & \cellcolor[HTML]{D7E219}\textcolor{black}{0.940} \\
Image FSIM & \cellcolor[HTML]{E4E318}\textcolor{black}{0.961} & \cellcolor[HTML]{E1E318}\textcolor{black}{0.954} && \cellcolor[HTML]{C5DF21}\textcolor{black}{0.911} & \cellcolor[HTML]{9FD938}\textcolor{black}{0.858} && \cellcolor[HTML]{E1E318}\textcolor{black}{\textbf{0.957}} & \cellcolor[HTML]{AADB32}\textcolor{black}{\underline{0.872}} && \cellcolor[HTML]{DFE318}\textcolor{black}{0.950} & \cellcolor[HTML]{E1E318}\textcolor{black}{0.957} && \cellcolor[HTML]{BFDF24}\textcolor{black}{0.906} & \cellcolor[HTML]{ADDC30}\textcolor{black}{0.876} && \cellcolor[HTML]{F1E51C}\textcolor{black}{0.980} & \cellcolor[HTML]{DCE218}\textcolor{black}{0.947} \\
Image NLPD & \cellcolor[HTML]{DFE318}\textcolor{black}{0.950} & \cellcolor[HTML]{DAE218}\textcolor{black}{0.945} && \cellcolor[HTML]{CDE01D}\textcolor{black}{0.922} & \cellcolor[HTML]{ADDC30}\textcolor{black}{0.879} && \cellcolor[HTML]{D4E11A}\textcolor{black}{0.936} & \cellcolor[HTML]{9AD83C}\textcolor{black}{0.851} && \cellcolor[HTML]{D4E11A}\textcolor{black}{0.936} & \cellcolor[HTML]{DAE218}\textcolor{black}{0.945} && \cellcolor[HTML]{C2DF22}\textcolor{black}{0.910} & \cellcolor[HTML]{B2DD2C}\textcolor{black}{0.884} && \cellcolor[HTML]{EEE51B}\textcolor{black}{0.973} & \cellcolor[HTML]{CDE01D}\textcolor{black}{0.923} \\
Image PSNR HVS & \cellcolor[HTML]{DFE318}\textcolor{black}{0.949} & \cellcolor[HTML]{DCE218}\textcolor{black}{0.945} && \cellcolor[HTML]{C2DF22}\textcolor{black}{0.909} & \cellcolor[HTML]{A7DB33}\textcolor{black}{0.871} && \cellcolor[HTML]{DAE218}\textcolor{black}{0.943} & \cellcolor[HTML]{97D83E}\textcolor{black}{0.845} && \cellcolor[HTML]{D2E11B}\textcolor{black}{0.932} & \cellcolor[HTML]{D7E219}\textcolor{black}{0.938} && \cellcolor[HTML]{C2DF22}\textcolor{black}{0.910} & \cellcolor[HTML]{B7DD29}\textcolor{black}{0.892} && \cellcolor[HTML]{EEE51B}\textcolor{black}{0.976} & \cellcolor[HTML]{CAE01E}\textcolor{black}{0.921} \\\bottomrule
\end{tabular}
\label{tab:per_content_correlation}
}
\end{table*}

Although the performance indexes computed across the entire dataset provide a good indication of both \textbf{cross-content} and \textbf{cross-codec} generalization, there are scenarios where requirements may be lower. 
For instance, when a metric is used to estimate the quality of point clouds compressed with a known codec, only \textbf{cross-content} generalization could be desired, %correct estimation of the quality no matter the rate nor the characteristics of the input model. 
which can be better evaluated by the six right-most columns of Table \ref{tab:entire_dataset_correlation}. % where the correlation was computed after splitting the dataset according to the coding engine. 
It is observed that correlation values with G-PCC are usually higher than for the other codecs, with many metrics achieving PLCC higher than 0.9 and significantly better performance than for the entire dataset. 
A possible explanation is the simplicity of octree-based geometric artifacts, as stronger compression leads to fewer points in the decoded point cloud and sparser distribution. 
GraphSIM achieves the best performance in this case, followed by MS-GraphSIM and image-based VMAF. 

\begin{figure}
    \centering
    \subfloat[Reference]{
        \begin{minipage}{0.3\linewidth}
            \includegraphics[width=\linewidth]{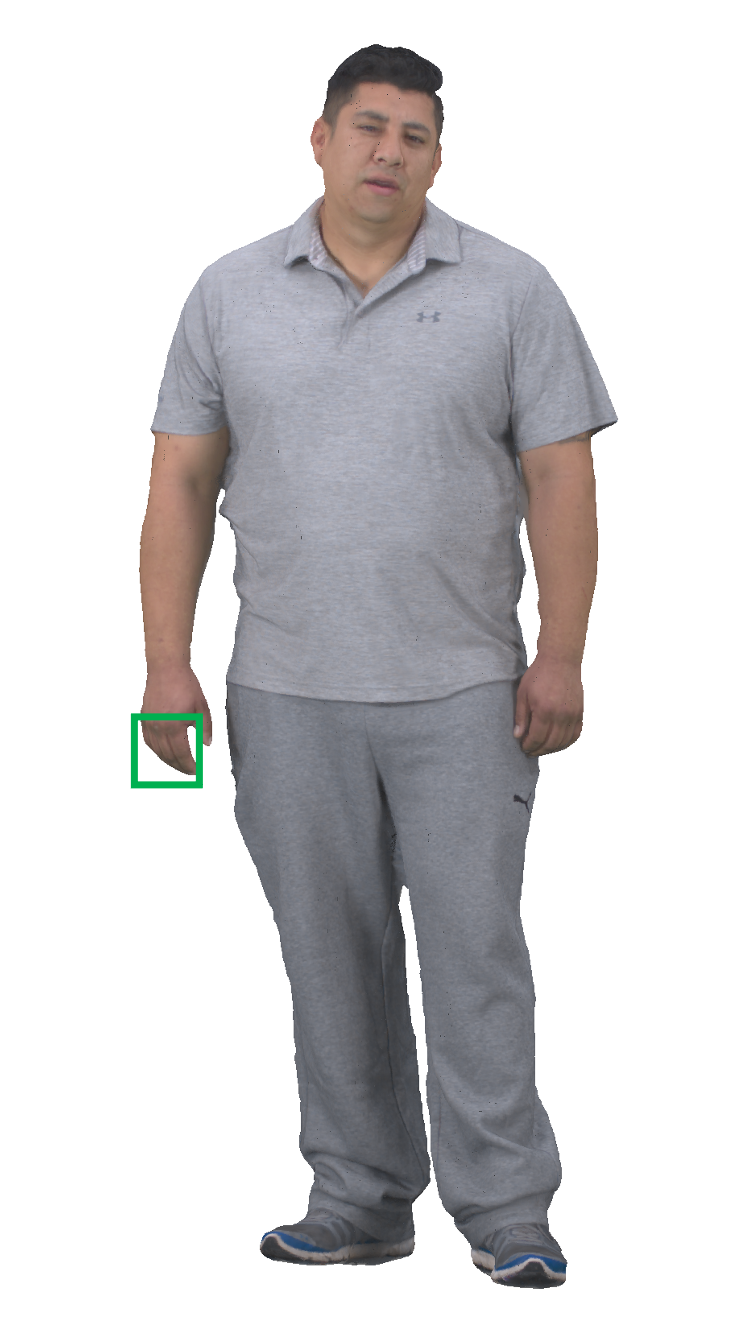}
            
        \end{minipage}
    }
    \subfloat[V-PCC R1]{
        \begin{minipage}{0.3\linewidth}
            \includegraphics[width=\linewidth]{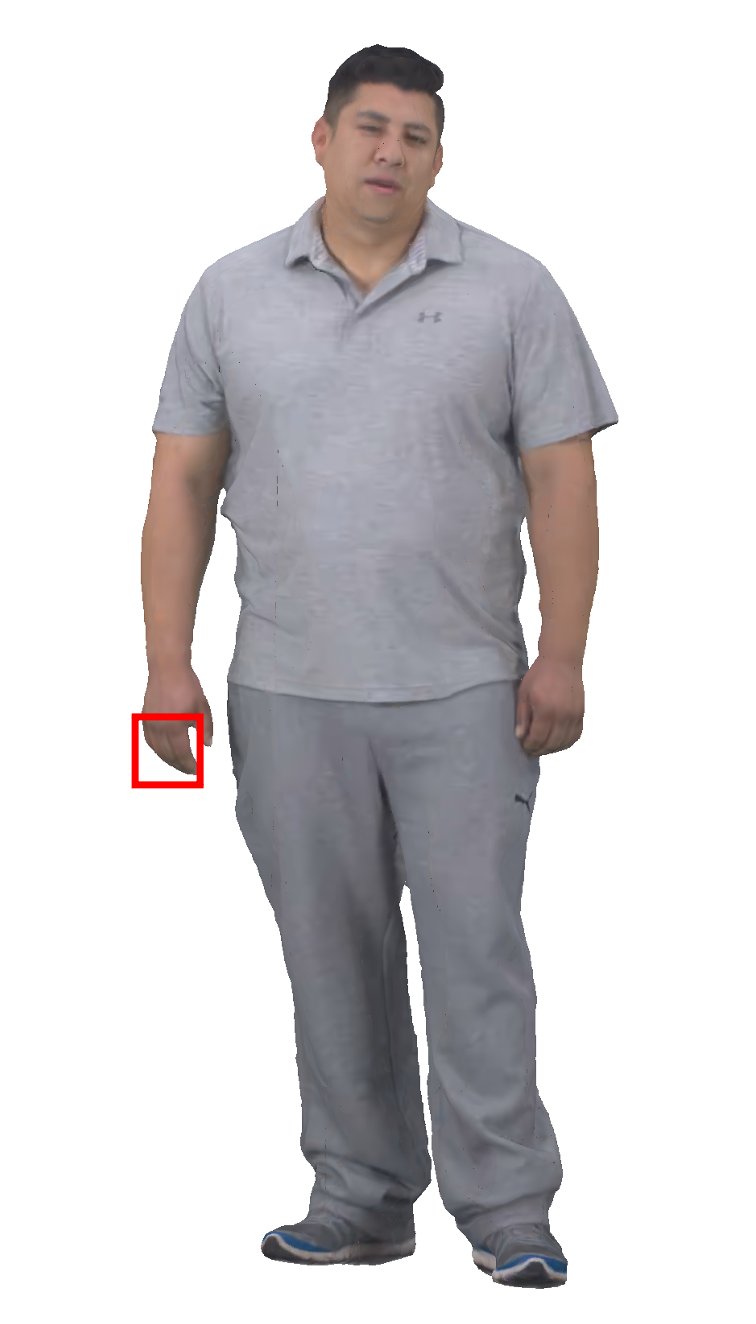}
        \end{minipage}
        }
    \begin{minipage}{0.3\linewidth}

        \fboxsep=0mm%padding thickness
        \fboxrule=0.8pt%border thickness

        \definecolor{darkgreen}{HTML}{00B050}
        \centering
        \fcolorbox{darkgreen}{white}{\includegraphics[scale=0.025]{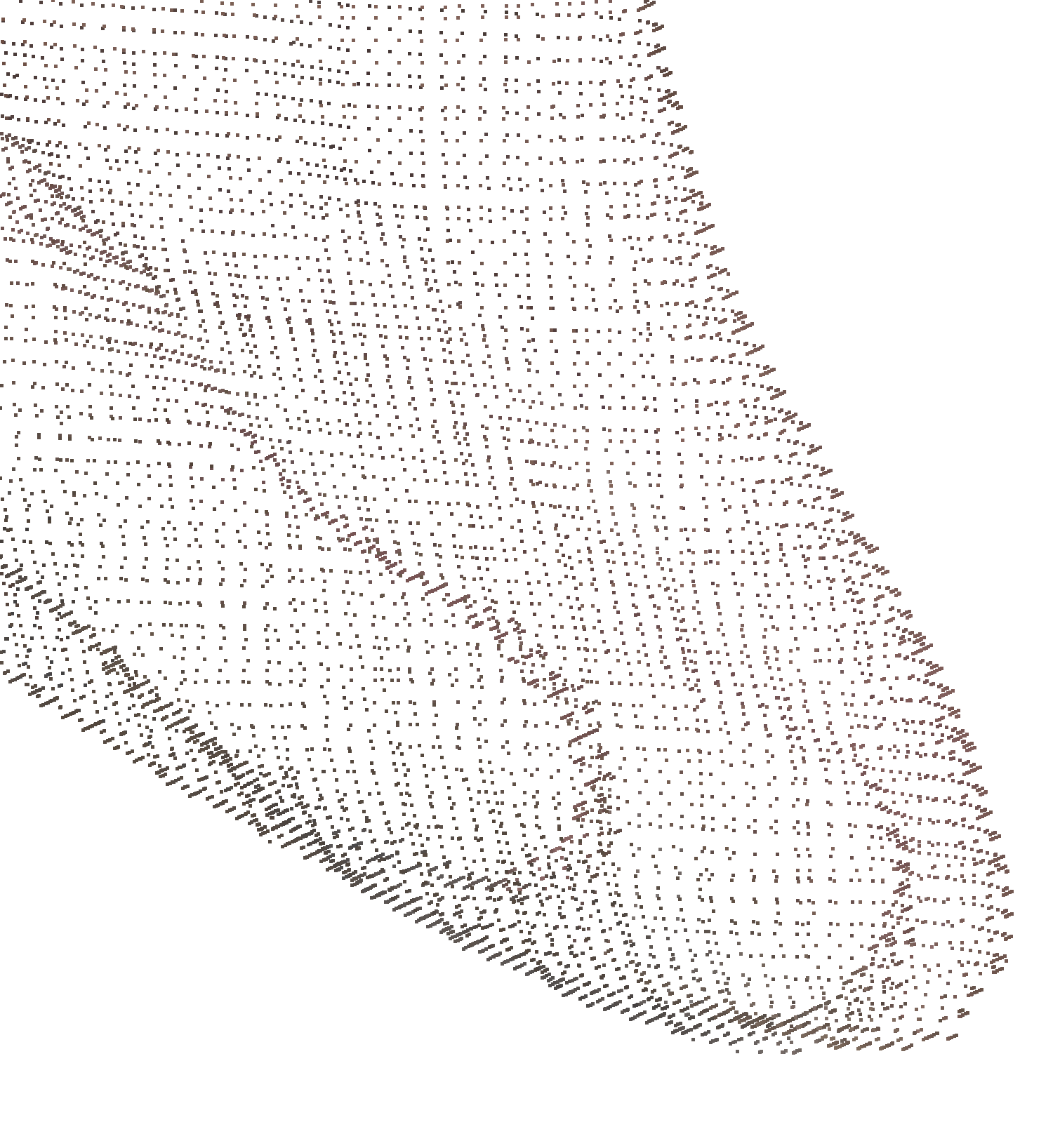}}
        \fcolorbox{red}{white}{\includegraphics[scale=0.025]{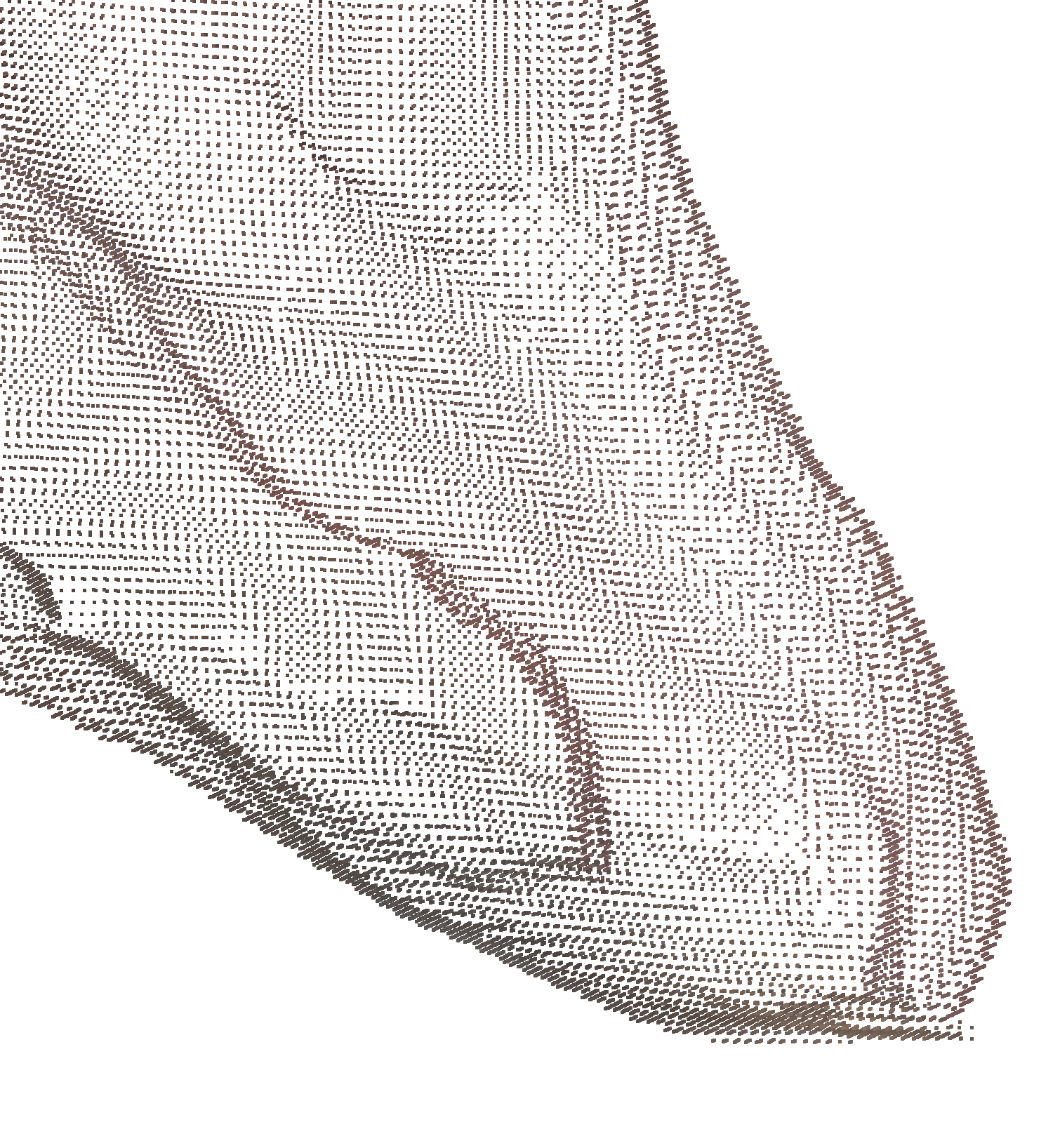}}

        \vspace{4pt}
        \fboxsep=1mm%padding thickness
        \fboxrule=0.8pt%border thickness
        \fcolorbox{black}{white}{\resizebox{0.75\linewidth}{!}{
            \parbox{\linewidth}{
            \centering
            MOS=4.75 
            }
            }
            }
        \vspace{1pt}
        \fcolorbox{black}{white}{\resizebox{0.75\linewidth}{!}{
            \parbox{\linewidth}{
            \centering
            PMOS: \\
            PCQM=4.31 \\
            GraphSIM=3.55 \\
            PointSSIM=2.99 \\
            }
            }            
        }
        
    \end{minipage}
    
    \caption{Original \emph{Boxer} and compressed with V-PCC at P1 and R1 displayed side-by-side, with a zoom on the hand region on the right. MOS value, as well as PMOS as given by Equation \ref{eq:logistic} for different metrics are also reported.}
    \label{fig:boxer_subjective}
\end{figure}

However, several metrics were not able to maintain high performance for V-PCC, where sharp drops in performance are observed for both point-to-distribution variants, Histogram Y and YUV, MS-GraphSIM, GraphSIM, and PointSSIM. 
For the last two metrics, it can be observed in Figure \ref{fig:scatter_plots} that stimuli generated from \emph{Boxer} and \emph{House\_without\_roof} with low objective scores received a high MOS. %  lower score than they should. 
As it can be seen in Figure \ref{fig:boxer_subjective}, V-PCC is able to keep relatively good quality even at the lowest rate, resulting in a high MOS value. 
However, the zoom on the hand of the model shows that the codec increases the point density, which cannot be perceived by subjects in the renderings displayed in the experiment but can affect the metrics score. 
Predictors computing features over neighborhoods defined with k-Nearest Neighbors such as GraphSIM and PointSSIM can be particularly affected since the neighborhoods will span over regions with smaller sizes for the denser model, potentially changing the distribution of the attributes. 
Image-based FSIM and PCQM achieve the highest performance, followed by MS-PointSSIM in terms of PLCC and MM-PCQA according to the SROCC. 
For JPEG Pleno, the maximum correlation values are in between G-PCC and V-PCC, with GraphSIM and LogP2D-JGY achieving the highest correlation. 
The performance of many metrics is more similar for JPEG Pleno and G-PCC than for V-PCC. 
For instance, Y PSNR and YUV PSNR are among the best-performing color-based metrics for V-PCC while the opposite is true for the other codecs. 
Likewise, several metrics that fail for V-PCC with PLCC under 0.4 have much better behavior for JPEG Pleno and G-PCC. 

% Some applications prioritize \textbf{cross-codec} generalization, especially when comparing the performance between compression standards. 
% Since rate-distortion curves are usually compared separately for each point cloud, \textbf{cross-content} generalization is less important. 
The recent rise in research of point cloud compression techniques providing high subjective quality highlighted the need for metrics with high \textbf{cross-codec} generalization ability since the performance of new codecs is evaluated by comparison of its rate-distortion curve to anchor coding methods separately for different point clouds. % as illustrated in Table \ref{tab:per_content_correlation}. 
Table \ref{tab:per_content_correlation} illustrates the PLCC and SROCC computed separately for each point cloud, which are indicators of \textbf{cross-codec} generalization. %, where most objective metrics achieve higher correlation when compared to previous evaluations. 
Most objective metrics achieve higher correlation when compared to previous evaluations and the highest coefficients are achieved for \emph{Bouquet}, \emph{Soldier}, and \emph{Thaidancer}, which are solid point clouds, i.e. with very high point density. 
% Even if the latter is voxelized with higher precision than the others, that 
The higher voxelization precision used for \emph{Thaidancer} of 12 bits does not seem to negatively impact the performance of the quality metrics compared to the other two point clouds where 10 bits were used. 
For these three models, FR metrics based either only on color features or both on geometry and color features usually achieve higher performance, with image-based metrics also having a very high correlation. 
For \emph{StMichael}, correlation coefficients are somehow lower, although this same set of metrics continues to correlate well with human perception. 

However, for the sparser point clouds \emph{Boxer} and \emph{House\_without\_roof}, the performance of point cloud metrics drops, with image-based metrics achieving the highest correlation coefficients. 
This effect is likely due to the way that different codecs, especially V-PCC, handle the sparsity of point clouds, as previously discussed and illustrated in Figure \ref{fig:boxer_subjective}. %, since G-PCC always reduces the number of points at the same time that V-PCC and JPEG Pleno can dramatically increase their density as discussed in \cite{lazzarotto2024subjective}. 
Since the point size is adapted in the rendering framework to produce watertight surfaces, the impact of these different effects on subjective perception is probably easier to capture directly in the projections than in the point domain. 
It is also observed that generally, geometry-only metrics perform worse than other FR metrics. 
While this effect is expected due to their lack of awareness about color distortion, the performance penalty for even the best geometry-based predictor is observed to be higher for \textbf{cross-codec} generalization than for \textbf{cross-content} generalization.

\section{Conclusions}

In this paper, a comprehensive set of objective quality metrics is assessed to obtain better insights into their correlation with subjective scores obtained in a point cloud quality experiment including compression artifacts generated by JPEG Pleno, G-PCC, and V-PCC. 
The evaluation is first conducted for the entire dataset where image-based metrics FSIM and VMAF were able to achieve the highest performance, with point cloud quality metrics PCQM, MS-PointSSIM, and GraphSIM coming next. 
The metrics were also separately assessed for cross-content and cross-codec generalization, showing that higher correlation coefficients can usually be achieved across different types of artifacts than across different point clouds. %for the same point cloud rather than the other way around. 
A drop in performance is observed when only sparser point clouds are considered, suggesting that current point cloud metrics are better suited to estimate visual quality for solid models. %are still not able to properly model different types of artifacts at wider ranges of point density. 
Future work may consider the insights provided in this study to develop more robust quality metrics, especially given the importance of quality predictors for the development and improvement of point cloud compression solutions.

%\section*{References}
\bibliographystyle{IEEEtran}

\end{document}